# Versatile coating platform for metal oxide nanoparticles: applications to materials and biological science

**Jean-François Berret[a]* and Alain Graillot[b]**

[a]Université Paris Cité, CNRS, Matière et Systèmes Complexes, Paris, France.
[b]Specific Polymers, ZAC Via Domitia, 150 Avenue des Cocardières, 34160 Castries, France.

**Abstract**: In this feature article, we provide an overview of our research on statistical copolymers as a coating material for metal oxide nanoparticles and surfaces. These copolymers contain functional groups enabling non-covalent binding to oxide surfaces and poly(ethylene glycol) (PEG) polymers for colloidal stability and stealthiness. The functional groups are organic derivatives of phosphorous acid compounds R-$H_2PO_3$, also known as phosphonic acids that have been screened for their strong affinity to metals and their ability to build multidentate binding. Herein we develop a polymer-based coating platform that shares features with the techniques of self-assembled monolayers (SAM) and Layer-by-Layer (L-b-L) deposition. The milestones of this endeavor are the synthesis of PEG-based copolymers containing multiple phosphonic acid groups, the implementation of simple protocols combining versatility with high particle production yields and the experimental demonstration of the colloidal stability of the coated particles. As an demonstration, coating studies are conducted on cerium ($CeO_2$), iron ($\gamma$-$Fe_2O_3$), aluminum ($Al_2O_3$) and titanium ($TiO_2$) oxides of different sizes and morphologies. We finally discuss applications in the domain of nanomaterials and nanomedicine. We evaluate the beneficial effects of coating on redispersible nanopowders, contrast agents for In Vitro/Vivo assays and stimuli-responsive particles.

Corresponding author: jean-francois.berret@u-paris.fr

## ■ INTRODUCTION

Coating is a field of surface science that aims at improving the properties of micro/macroscopic surfaces. There is currently a wide variety of physical and chemical coating techniques consisting in altering the atoms and molecules at interfaces or in overcoating the existing surface with materials of the desired properties.[1] Depending on applications, the coating thickness range from a few nanometers for ultrathin molecular films to micrometers for thick films. With the advent of nanotechnology and the reduction of material size in applications, coating strategies have evolved to accommodate the miniaturization of substrates. One area where coating has become important is that of nanoparticles (NPs). NPs are objects with at least one dimension comprised between 1 and 100 nm. The NPs are synthesized in an aqueous or organic solvent, under controlled physicochemical conditions that ensure their stability during the synthesis reaction. However, coating NPs dispersed in a solvent raises specific issues and the coating techniques for micro/macroscopic surfaces must be adapted accordingly. In particular, NPs have rotational and translational degrees of freedom that allow them to change the state of the dispersion *via* mechanisms such as aggregation or precipitation.[2]

As an illustration, we consider here two representative examples, one in the field of nanomaterial chemistry and the other in nanomedicine. Metal oxide (MOx) nanoparticles[3] are typically synthesized in aqueous media by soft chemistry bottom-up approaches from metal salt solutions.[4-6] The nanocrystal growth is driven by homogeneous nucleation, and in many cases long-range electrostatic interactions are used to control the NP growth and ensure stability.[7] Studies have shown that changing the physicochemical conditions of cationic MOx dispersions, for instance by increasing the pH or ionic strength to bring them to physiological conditions leads to the destabilization of the dispersion and to the agregation and precipitation of the NPs.[8] In the realm of nanomedicine, NPs are put in contact with biological solvents, cells or tissues for imaging, diagnostic, therapy and engineering purposes. In these examples, they must be protected against the adsorption of endogenous biomolecules such as proteins, lipids, amino acids, nucleotides etc.. at their surface. The concept of protein corona[9-10] developed *circa* ten years ago encompasses the above cases, and it has been shown that this adsorption can deeply modify the synthetic identity of the NPs.[11] Recent surveys have shown that nanomedicine has indeed not met initial expectations in terms of clinical translation. Based on the meta-analysis on more than 100 publications between 2005 and 2015, Wilhelm *et al.* have found that for the treatment of solid tumors in animal models, 99.7 % of administered particles, acting here drug vectors, were cleared from the blood circulation, and eventually miss their target, the solid tumor.[12] One of the reasons given for this setback was that the physicochemical interactions of NPs with biological environments have been overlooked and that NP interfacial properties In Vitro/Vivo have not been studied with sufficient emphasis.[12-13]

In practice, coating is necessary to make NPs compatible with the subsequent stages of their uses and transformations during their life cycle. The coating of NPs thus has several objectives that must all be achieved, by *i)* adjusting the interparticle interaction potential to the required physicochemical conditions, *ii)* preserving the size-related and surface properties of NPs, and *iii)* increasing the resistance to protein adsorption, a property known as anti-fouling.[14-16] For NP coating, the most common technique used to date is derived from the micro/macroscopic surface coating methods and consists in the chemical modification of the surface by adding an organic layer, or adlayer. This adlayer is made up of low molecular weight ligands, surfactants, phospholipids, oligomers or polymers of various architectures[17-18]. The addition of this organic layer is usually produced following a pathway described as *two-step coating process*. Here, NPs and coating molecules are synthesized separately (step 1) and later assembled following appropriate protocols (step 2). This two-step pathway actually differs from another widely used process where the organic (macro)molecules are introduced during synthesis to control the nanocrystal nucleation and growth as well as to ensure colloidal stability. The advantages of the *two-step coating process* are numerous and have recently led to remarkable advances. In particular, this process combines simplicity, versatility and high particle production yields. Another advantage is that it encompasses a wide range of chemical techniques, such as the *grafting-from* and *grafting-to* techniques. The latter strategy also includes ligand adsorption, polymer grafting, self-assembled monolayers, layer-by-layer deposition, and phase transfer of coatings between organic and aqueous solvents.[17-23] In this review, we will focus on the application of the t*wo-step coating process* to the class of metal oxide nanoparticles (MOx-NPs), which represent a large collection of nanostructures currently under intense research. The emphasis will be put on cerium and iron oxide NPs ($CeO_2$ and $\gamma$-$Fe_2O_3$ respectively), and we will give examples of efficient coatings for titanium ($TiO_2$) and aluminum ($Al_2O_3$) oxides. The interest in MOx-NPs comes from the combination of complementary attributes, such as nanometer size, high surface reactivity and remarkable properties in catalysis, electronics, optics and magnetism. We refer the reader to the following reviews for the applications of MOx-NPs.[4,24]

In the field of NP coating, macromolecules of synthetic and biological origin have become the most readily used species for surface modification. There are a number of underlying reasons to this: *i)* there exists a large library of polymer architectures including linear chains, copolymers, stars, dendrimers which can be implemented as coating; *ii)* the theoretical framework of polymers at interfaces is well founded and the predictions have been validated by experiments for more than 2 decades;[25-27] *iii)* polymers are relatively inexpensive to produce and some have been approved by regulatory and control agencies for medical applications. As for polymers conformation at interfaces, Alexander and de Gennes provided the theoretical scheme for polymers tethered by one extremity and having the other extremity dangling in the solvent.[28-29] On flat surfaces, two main regimes were predicted. At polymer density $\sigma$ such as $\sigma < 1/\pi R_g^2$ (i.e. typically $\sigma$ < 0.1 nm$^{-2}$), where $R_g$ is the gyration radius of the chain in good solvent, the polymers adopt a mushroom configuration, the adlayer thickness being then twice the gyration radius. At higher densities, $\sigma > 1/\pi R_g^2$ ($\sigma$ > 0.1 nm$^{-2}$), monomer-monomer excluded volume interactions induce a stretching of the chains, which enter into the brush regime. In this configuration, the height increases and varies as $h_{2D} \sim \sigma^{\nu} N$, where $N$ is with the degree of polymerization and $\nu$ a coefficient between 1/3 and 1 depending on the solvent quality.[27,30] Polymer adlayers or brushes have been shown to form soft interfaces and efficient steric repulsive barriers.[14]

Polymer affinity towards surfaces can be enhanced by the addition of specific chemical groups that can react with the surface. For MOx-NPs, the most commonly used linkers are alcohol, acid, amine, silane and thiol compounds.[3,17] In the above list, organic derivatives of phosphorous acid with the formula R-$H_2PO_3$, also referred to as phosphonic acids, have attracted much attention. Since the pioneer work by Schwarzenbach and coworkers,[31] phosphonic acid is known to have a strong affinity toward metallic and metal oxide surfaces compared to sulfates and carboxylates.[3,32-34] The binding mechanisms were identified as resulting from condensation of acidic P-OH hydroxyls with surface metal hydroxyls, or from coordination of phosphoryl oxygen to Lewis acid surface sites.[3,32] The prevalence of one mechanism over the other depends on the reaction conditions, and on the type of oxide. Moreover, the presence of three oxygen atoms on phosphonic acid allows multidentate binding modes,[32,35] in combination with possible hydrogen-bonding interactions (**Fig. 1**), as shown by advanced surface characterization techniques based on NMR[32,35] and mass spectrometry.[36] Some authors have also suggested that the three oxygen atoms could bind to the same metal site, or to different metal atoms on the surface.[3] Previous reports on MOx functionalization have shown that PEG polymers terminated with mono or bi-phosphonic acid functional groups can adsorb on particles of different composition and nature, such as calcium carbonate,[33] iron,[37-40] cerium[41-42], titanium[43-44] and zinc-gallium[45] oxides. More recently, functional polymers with multiple phosphonic acid groups have been shown to outperform polymers with a single functional group in terms of resilient coating.[43,46]

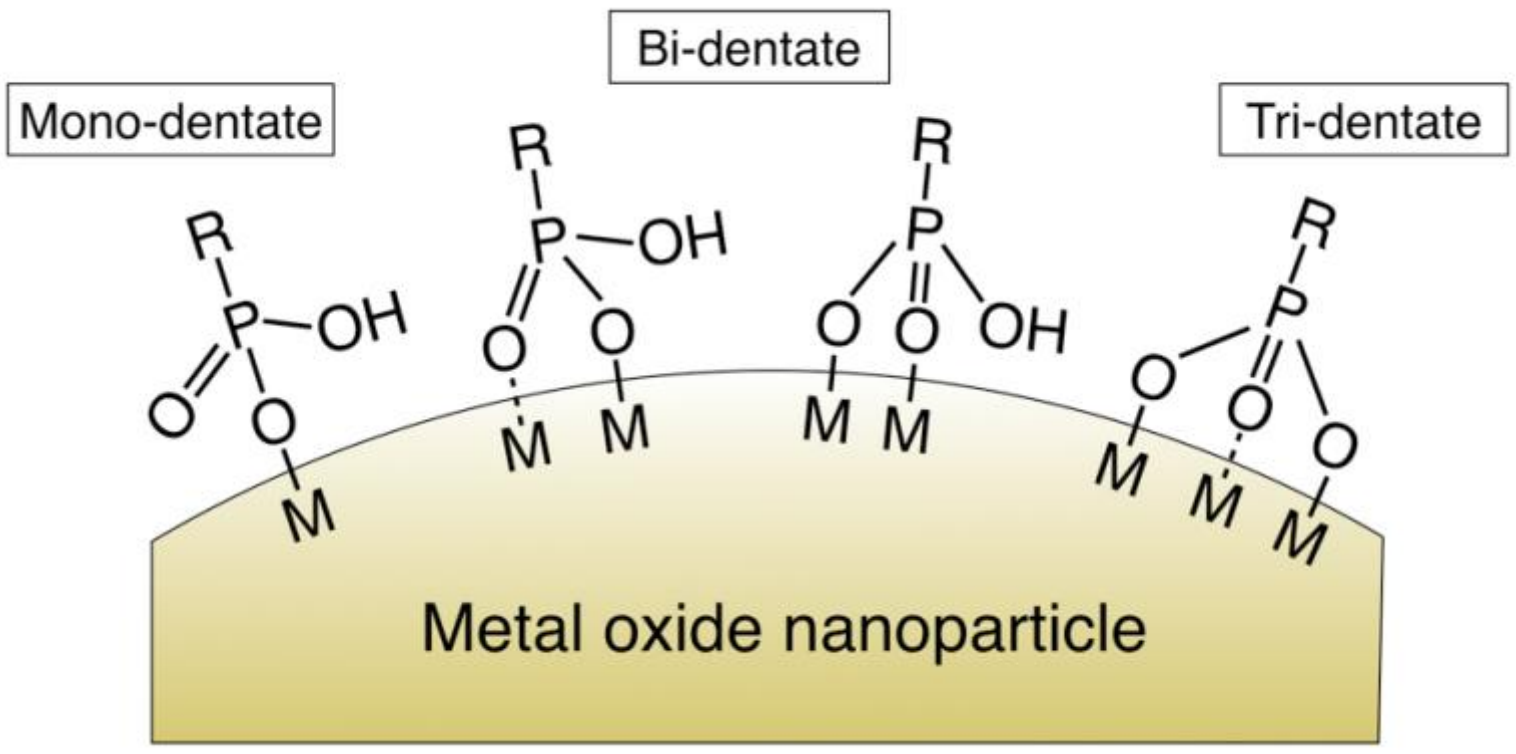

**Figure 1:** Schematic representation of mono-, bi- and tri-dentate binding modes of phosphonic acids upon adsorption on metal oxide nanoparticle surface.[32,47-48]

Here we review the course of research that led to the development and application of functional polymers containing phosphonic acids as effective coating material for metal oxide nanoparticles. The milestones of this endeavor are the synthesis of PEG-based copolymers containing multiple phosphonic acid groups (R-$H_2PO_3$), the implementation of simple protocols combining versatility and high particle production yields and the experimental demonstration of the coated particle colloidal stability. The chosen approach is based on the aforementioned *two-step coating process* which allows to evaluate the intrinsic performance of a coating adlayer as an application, whether in the field of materials or nanomedicine.

## ■ PHASE BEHAVIOR AND COLLOIDAL STABILITY OF COATED METAL OXIDE NANOPARTICLES

In this review, we will adopt the following notations. Oligomers or polymers terminated by a single acidic phosphonic acid will be referred to as $OEG_{ik}$-Ph or $PEG_{ik}$-Ph (**Fig. 2a**). For chains with multiple phosphonic acids, the abbreviation $MPEG_{ik}$-*co*-MPh will be used to describe the general class of statistical copolymers investigated. There, $MPEG_{ik}$ denotes a PEG methacrylate monomer with a PEG molecular weight of i kDa, whereas MPh stands for a methacrylate monomer bearing a phosphonic acid functional group (**Fig. 2b**). In a similar way, for copolymers bearing carboxylic acid groups instead of phosphonic acid moieties will be called $MPEG_{ik}$-*co*-MA (**Fig. 2c**). For a specific copolymer chain with a known number *n* of phosphonic acid or methacrylic acid functions, $(MPEG_{ik})_n$-*co*-$MPh_n$ or $(MPEG_{ik})_n$-*co*-$MA_n$ will be used. In this review we will also highlight more complex chemical structures, such as $MPEG_{ik}$-*co*-$MPEG_{jk}$a-*co*-MPh terpolymers containing an amine-terminated PEG chain allowing chemical substitution (**Fig. 2d**), and MPAm-*co*-MPh copolymers made from a methacrylate monomer bearing a propylacrylamide group (**Fig. 2e**).

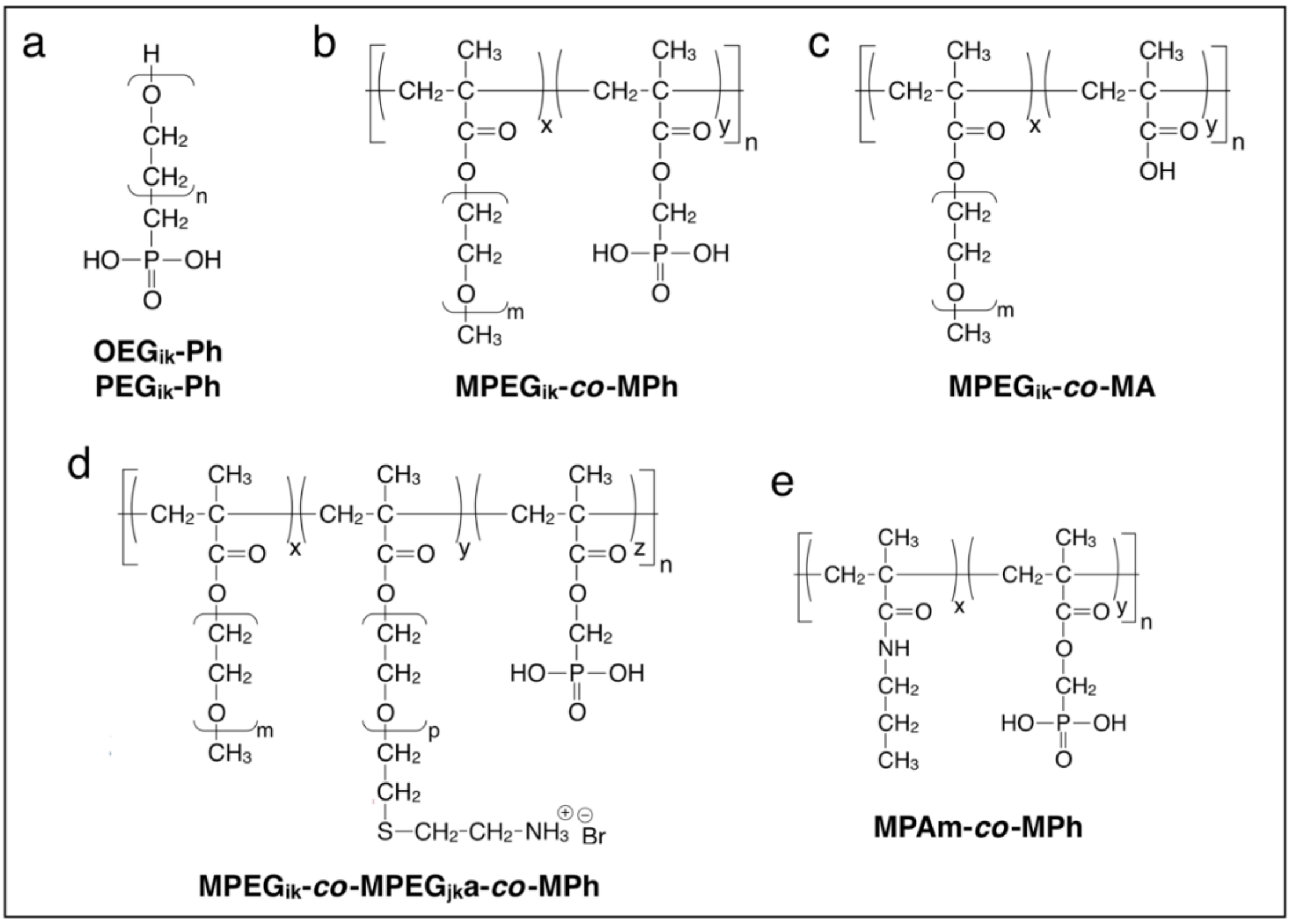

**Figure 2**: Chemical structures and abbreviations of the polymers studied in this work. **a)** Oligo(ethylene glycol) (i < 0.6) and poly(ethylene glycol) (i > 0.6) of molecular weight i kDa[1] terminated with a single phosphonic acid. **b)** Statistical copolymer $MPEG_{ik}$-*co*-MPh made from $MPEG_{ik}$ and MPh monomers. **c)** Statistical copolymer $MPEG_{ik}$-*co*-MA made from $MPEG_{ik}$ and MA monomers. **d)** Statistical terpolymer $MPEG_{ik}$-*co*-$MPEG_{jk}$a-*co*-MPh made from $MPEG_{ik}$, $MPEG_{jk}$a and MPh monomers, where $MPEG_{ik}$a is an amine-terminated $PEG_{ik}$ methacrylate. **e)** Statistical copolymer MPAm-*co*-MPh made from MPAm and MPh monomers, where MPAm is a methacrylate monomer bearing a propylacrylamide group.

## From PEGylated coating to redispersible nanopowders

Our work on functional macromolecules for coating was initiated by a collaboration with Solvay® regarding the development of redispersible nanopowders.[49] From an industrial perspective, nanopowders represent an economically interesting state of matter, because of their ease of transport or handling, their increased shelf life and the possibility to formulate dispersions with various solvents. Studies have shown that when colloids are concentrated to the dry state, the van der Waals attraction tends to stick them together irreversibly,[2] a phenomenon that is further enhanced with MOx-NPs because of their high specific surface area and crystalline facets. For this first study, we considered cerium oxide ($CeO_2$) nanoparticles (**Fig. 3a**) that are used in industrial applications as catalysts in the automotive and chemical-mechanical polishing sectors, and as UV absorbers in paint formulations.[50-51] The coating molecule investigated, a phosphonic acid terminated oligo(ethylene glycol) of molecular weight 0.4 kDa ($OEG_{0.4k}$-Ph, **Fig. 3b**) is Solvay proprietary oligomer containing 10 ethylene glycol monomers. In search for protocols with high yields and upscaling potential, the *two-step coating process* was implemented. With this technique, the oligomers and NPs were synthesized separately and later assembled according to a protocol described in details below. Synthesized by thermohydrolysis of cerium salt at pH 1.4, the 7.8 nm $CeO_2$-NPs shown in **Fig. 3a** carry positive charges on their surface, ensuring their colloidal stability.[52-53] However, a small increase in pH, e.g. of the order of one pH-unit leads to their destabilization and aggregation.[54] For these reasons, we suggested that the coating protocol must be carried out under pH and salinity conditions identical to that of as-synthesized $CeO_2$ batches.[41,54]

To study the phase behavior of $CeO_2$ and $OEG_{0.4k}$-Ph, we combine the electrostatic complexation effect between the cationic NPs and anionic oligomers with the continuous variation method developed by P. Job.[55-59] $CeO_2$-NPs and oligomers were prepared in the same conditions of pH (1.4) and concentration, and mixed at different volumetric ratios $X$. The continuous variation method allows exploring a broad range of conditions, while keeping the total concentration in the dilute range. $X$ is varied from $10^{-3}$ to $10^3$, switching from a regime of excess oligomers ($X \ll 1$) to a regime of excess NPs ($X \gg 1$). Upon mixing the oligomers spontaneously adsorbed on the particle surface *via* electrostatic complexation, and form a 1.8 ± 0.2 nm layer, as shown by small-angle scattering.[41] The different steps of the protocol are illustrated in **Fig. 3c.**

**Fig. 3d** displays the hydrodynamic diameter map of $CeO_2$@$OEG_{0.4k}$-Ph in the ($X$ - pH) diagram for $X$ varying from 0.1 to 10 and pH from 1 to 14. The region in red shows the domain of stability of the coated particles, whereas the shaded area displays the precipitated region. The major finding of this study is the existence of a critical ratio $X_C$, here equal to 1 below which the coated NPs are stable over a broad pH range (1 < pH < 9). For $X > X_C$, the particles are aggregated or precipitated. This aggregation can be explained by the fact that the cationic sites $CeOH_2^+$ at the surface[53] are not saturated with oligomers, and that the OEG layer is not sufficiently dense to counteract the attractive van der Waals interaction and the electrostatic charge changes ($CeOH_2^+$ becoming CeOH with increasing pH) observed upon pH increase.[52,60] Light and neutron scattering data showed that at saturation there is an average of 270 $OEG_{0.4k}$-Ph oligomers per particle, corresponding to an oligomer density of $\sigma$ = 1.4 $nm^{-2}$ and a proportion of 40% by weight of the core-shell hybrid.[41] $CeO_2$ nanopowder were obtained by freeze-drying a mixed dispersion prepared with excess oligomers following by a dialysis step to remove the unreacted species. In contrast to bare $CeO_2$, the $CeO_2$@$OEG_{0.4k}$-Ph powder was found to redisperse readily in DI-water as well as in non water-borne solvents,[61] where dispersions exhibited non-aggregated NPs only. Interestingly, this approach allowed us to lay the basis for generic protocols that were later implemented on a wide variety of nanoparticles and macromolecular coating, with direct applications to the biomedical technologies.

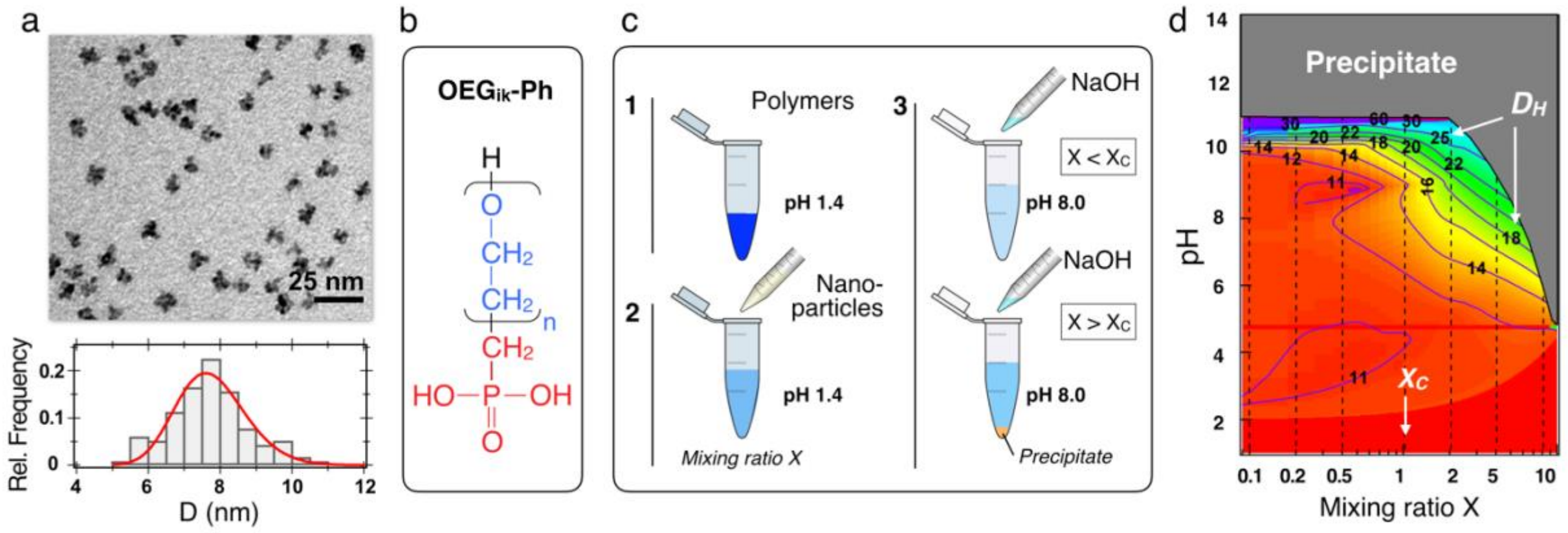

***Figure 3: a)** Upper panel: transmission electronic image of cerium oxide nanoparticles. Lower panel: Relative frequency as a function of TEM diameter. The continous line results from non-linear least squares analysis using a lognormal centered at 7.8 nm. **b)** 3-phosphonopropyl alcohol ethoxylate-10 EG, in short $OEG_{0.4k}$-Ph is an oligomer containing 10 ethylene glycol monomers and terminated by a phosphonic acid group. **c)** Coating protocol carried out for grafting $OEG_{0.4k}$-Ph at*

*$CeO_2$-NP surfaces. The association between anionic oligomers and cationic NPs is driven by electrostatic interaction at pH 1.4. At this pH, phopshonic acids have a deprotonation rate of 5%.[62] **d)** ($X$-pH) state diagram of $CeO_2$@$OEG_{0.4k}$-Ph nanoparticles obtained by dynamic light scattering at total concentration 10 g $L^{-1}$. $X$ denotes the concentration ratio between NPs and oligomers. The contour lines indicate hydrodynamic diameter $D_H$ between 10 to 60 nm, whereas the grey region represents phase separation. $X_C$ represents the critical mixing ratio above which NPs aggregate at physiological pH. Adapted with permission from Ref.[63] (Copyright 2020 American Chemical Society) and Ref.[41] (Copyright 2008 American Chemical Society).*

**Multiple phosphonic acid PEG copolymers: a robust coating platform**

Shortly after the work on redispersible nanopowders, the question of enhancing the adsorption energy of anchoring moities has been addressed, e.g. by changing the nature of the anchoring groups or increasing their numbers per chain. New PEGylated polymers with multidentate catechol[64-67], dopamine[67-71] or phosphonic acid[38-40,46-47,62,72-79] functional groups were evaluated and applied to various NPs and substrates. In this context, iron oxide nanoparticles have attracted attention and have been studied as contrast agents for magnetic resonance imaging (MRI) in biomedical applications.[80-82] Our interest was focused on sub-10 nm iron oxide NPs that we sought to coat with copolymers containing multiple phosphonic acids and PEGs per chain. The synthesis was performed by free radical polymerization by Specific Polymers®, with the objective to harness the protocols set up for redispersible nanopowders. The copolymers studied consist in alternating PEG chains and phosphonic acidic moieties grafted onto a poly(methyl methacrylate) backbone, as illustrated in **Fig. 4a**. For the copolymer noted $(MPEG_{2k})_3$-*co*-$MPh_3$ investigated in this section, the number of phosphonic acids and $PEG_{2k}$ chains was 3.1 in average, and its weight-average molecular weight and molar mass dispersity were 13.0 ± 0.5 kDa and 1.8. $(MPEG_{2k})_3$-*co*-$MPh_3$ coating agent was evaluated with 6.8 nm (**Fig. 4b**) and 13.2 nm iron oxide nanoparticles.[75]

**Figs. 4c** and **4d** displays the stability diagram of $\gamma$-$Fe_2O_3$@$(MPEG_{2k})_3$-*co*-$MPh_3$ mixed dispersions at the concentration of $c$ = 1 g $L^{-1}$ and pH 8.0. In **Figs. 4c**, images of vials containing the dispersions at different $X$-values are presented. The color change of the dispersions is due to the variations of $\gamma$-$Fe_2O_3$ concentration as $c_{NP} = cX/(1+X)$, whereas at the same time, the polymer concentration decreases as $c_{Pol} = c/(1+X)$. In **Figs. 4d**, the hydrodynamic diameter $D_H$ is displayed as a function of $X$. The hydrodynamic diameters of single polymer chains and $\gamma$-$Fe_2O_3$-NPs are 16 and 14 nm (squares at $X = 10^{-3}$ and $10^3$, respectively). Upon NP addition, $D_H$ exhibits first a plateau over 2 decades in $X$, and then increases sharply due to the formation of micron-sized aggregates. In this range, the dispersions are turbid and sediment with time. The existence of a critical mixing ratio $X_C$ is found again, which can be interpreted as follows. Below $X_C$, the polymers are in excess, the functional end-groups exceed the number of binding sites (here $FeOH_2^+$)[83] and after adsorption, coverage is maximum. Above $X_C$, the coverage is incomplete and the particles precipitate upon pH increase, as bare particles do. In addition to its simplicity, the previous protocol allows to determine the number of adsorbed chains per particle $n_{ads}$, *via* the relationship:[84]

$$n_{ads} = \frac{1}{X_C}\frac{M_n^{NP}}{M_n^{Pol}} \qquad (1)$$

where $M_n^{NP}$ and $M_n^{Pol}$ are the number-averaged molecular weights of the NP and polymer, respectively. For the 6.8 nm iron oxide particles, $X_C$ = 1.3 and $n_{ads}$ = 97, whereas for 13.2 nm NPs, $X_C$ = 5 and $n_{ads}$ = 230. These $n_{ads}$-values correspond to a copolymer density $\sigma$ = 0.50 ± 0.10 nm$^{-2}$ and to a $PEG_{2k}$ density $\sigma_{PEG}$ = 1.5 nm$^{-2}$, this latter value being regarded as high for a non-covalent grafting-to technique.[85-86] Comparing the hydrodynamic sizes of bare and coated NPs, the $PEG_{2k}$ thickness was evaluated at 6 nm, an outcome consistent with partially stretched chains.[14,87] Finally, electrokinetic measurements confirmed that the PEGylated particles were globally neutral. Last but not least, the previous protocol has been later upscaled to larger volumes, resulting in quantities of $\gamma$-$Fe_2O_3$@$(MPEG_{2k})_3$-*co*-$MPh_3$ of the order of 1 gram (dry extract) in a single coating protocol. This property has proven to be crucial for In Vitro/Vivo assays, which requires large quantities of particles.[78]

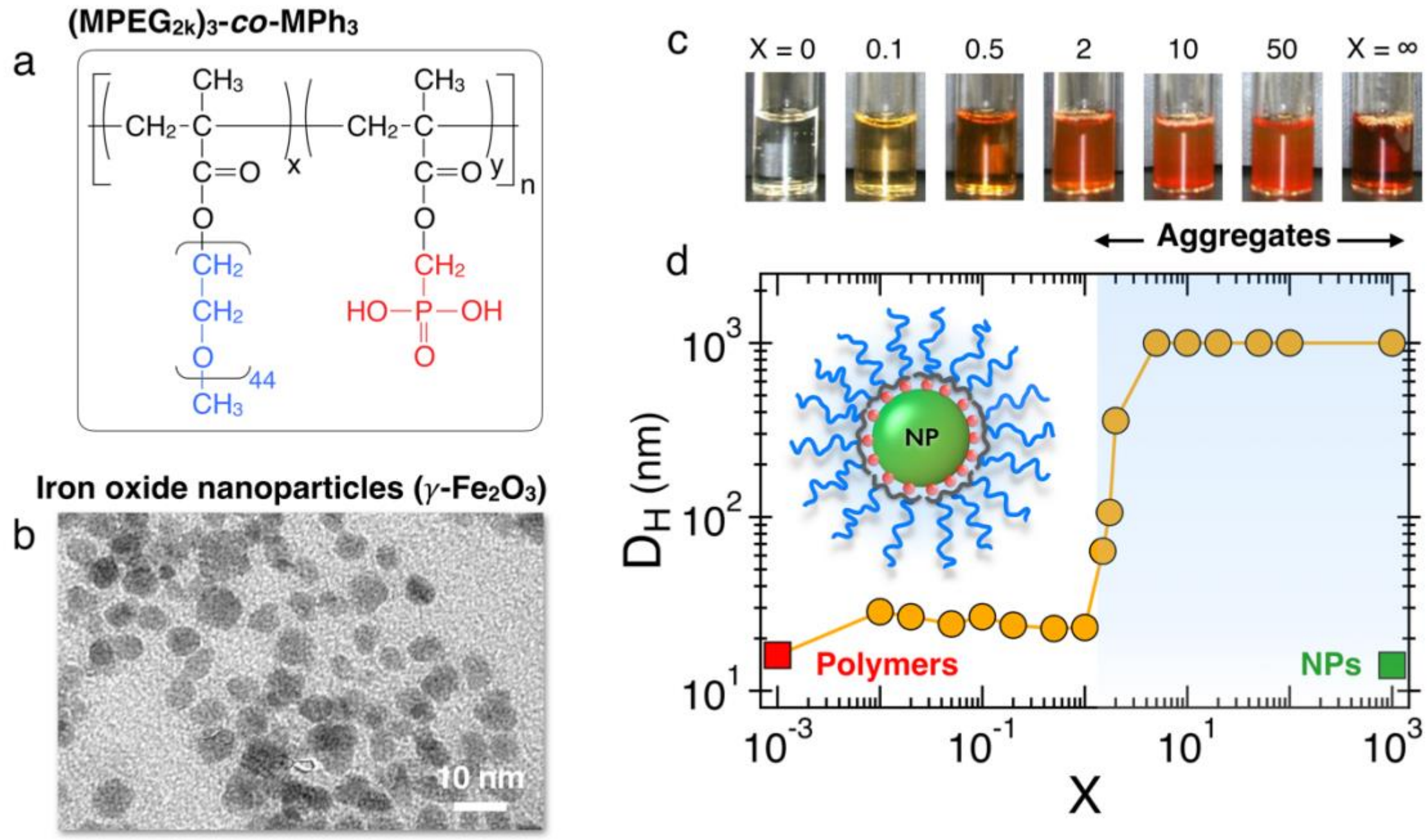

***Figure 4: a)*** *Chemical formula of the statistical copolymer* $(MPEG_{2k})_3$*-co-*$MPh_3$ *of weight-averaged molecular weight* $M_w$ *= 13.0 ± 0.5 kDa and molar mass dispersity Đ = 1.8 4. The comonomer proportions are* $x$ *=* $y$ *= 0.50, for a degree of polymerization of* $n$ *= 6.* ***b)*** *Transmission electron microscopy (TEM) of the 6.8 iron oxide nanoparticles.* ***c)*** *Images of vials containing* $\gamma$*-*$Fe_2O_3$*@*$(MPEG_{2k})_3$*-co-*$MPh_3$ *dispersions at pH 8.0 for different mixing ratio.* ***d)*** *Hydrodynamic diameter of* $\gamma$*-*$Fe_2O_3$*@*$(MPEG_{2k})_3$*-co-*$MPh_3$ *dispersions as a function of the mixing ratio* $X$ *(*$c$ *= 1 g L*$^{-1}$*, pH 8.0, T = 25 °C). The association between anionic copolymers and cationic NPs is driven by electrostatic interaction at pH 2.0. At this pH, phosphonic acids have a deprotonation rate of 17%*[62]*. Above* $X_C$ *= 1.3, dispersions are turbid due to aggregation. For large aggregates,* $D_H$ *is set at 1 µm. Inset: schematic representation of a coated NP. Adapted with permission from Ref.*[75] *(Copyright 2014 American Chemical Society).*

## Metal oxide nanoparticle stability in complex media

For application in materials or biomedical science, before any use, the colloidal stability of NPs must be assessed. This assessement is realized by examining the state of the dispersion as a function of time, over periods ranging from a few seconds to several days or weeks. Among the many techniques available for this purpose, static and dynamic light scattering is one of the most convenient.[8] A dispersion is considered stable if the scattered intensity $I_S(t)$ and the hydrodynamic diameter $D_H(t)$ obtained by light scattering are both stationary in time, and if $D_H$ is identical to its value in DI-water. As $I_S$ is proportional to the molecular weight of the scatterers,[88] its monitoring is more sensitive than that of $D_H$ to detect slow or partial aggregation processes.[89] For stability studies, it is recommanded to examine solvents of increasing complexity. These solvents includes brine (e.g. DI-water at 1 M NaCl), phosphate buffer saline (PBS), cell culture medium with and without serum, and plasma serum. Typical cellular media are the Dulbecco's Modified Eagle's medium, DMEM and the Roswell Park Memorial Institute medium, RPMI, which both contain inorganic salts, amnio acids, vitamins and other components such as glucose. As for the protocol, a few microliters of a concentrated dispersion (e.g. $c$ = 20 g $L^{-1}$ in DI-water) are poured and homogenized rapidly in 1 mL of the solvent to be studied, and simultaneously $I_S$ and $D_H$ are measured over 2 hours and later at day 1, day 7 and day 30. The end concentration is chosen according to the concentrations used in In Vivo/Vitro assays and to the sensitivity of the light scattering spectomerter.

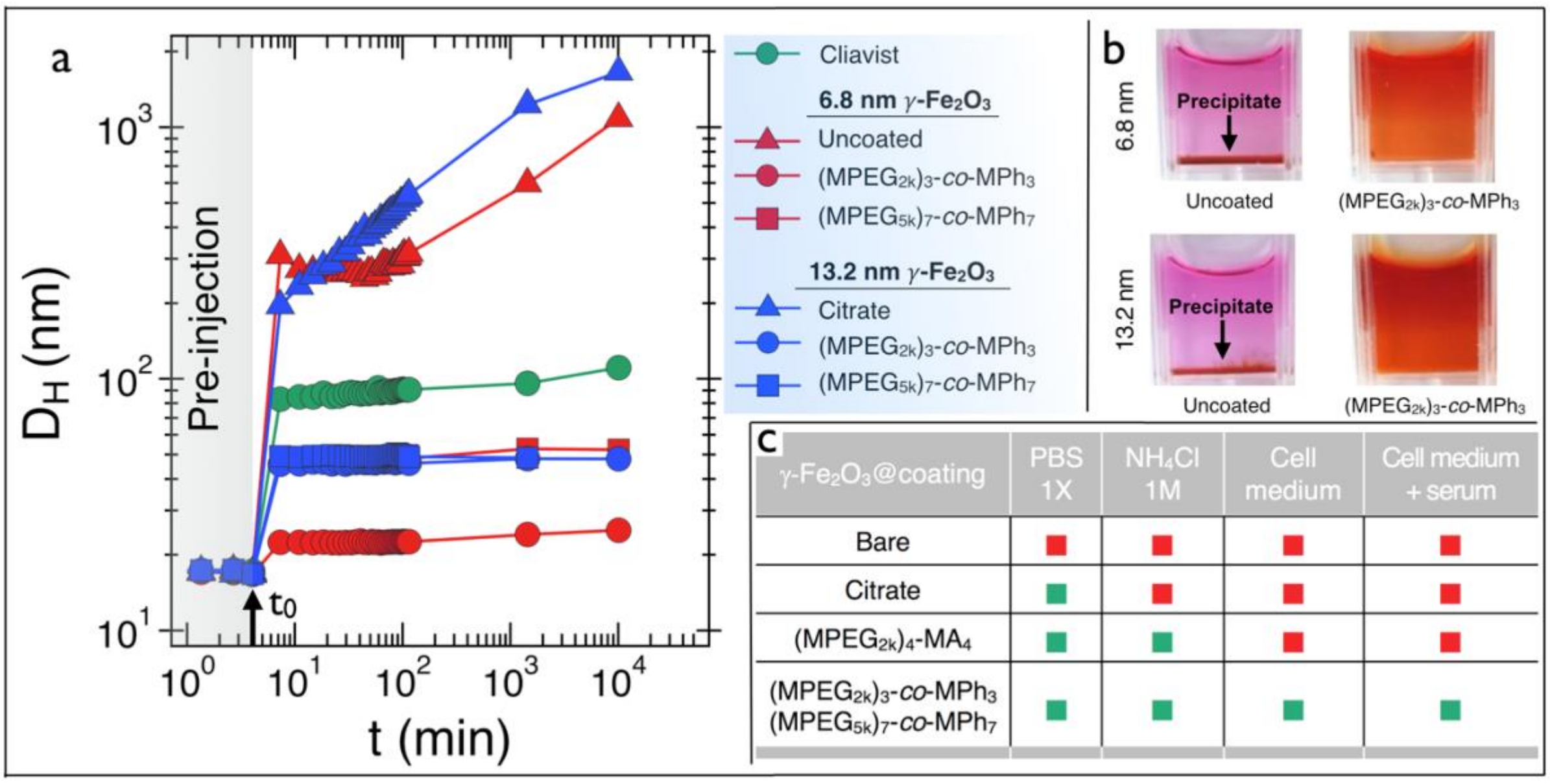

***Figure 5: a)*** *Hydrodynamic diameter $D_H$ versus time for γ-$Fe_2O_3$@$(MPEG_{2k})_3$-co-$MPh_3$ and γ-$Fe_2O_3$@$(MPEG_{5k})_7$-co-$MPh_7$ after their injection at $t_0$ in Dulbecco's Modified Eagle's Medium with serum. Also included are the time behavior of uncoated γ-$Fe_2O_3$ NPs and of commecial contrast agent cliavist®, Bayer (Germany). Cliavist® contains partialy aggregated 4 nm magnetic core particles coated with 1400 Da carboxydextran polymers[82] and was developed for liver imaging using MRI.* ***b)*** *Images of iron oxide dispersions in DMEM after one week.* ***c)*** *Comparison of the colloidal stability of bare and coated 13.2 nm γ-$Fe_2O_3$ nanoparticles in various complex media. Green squares indicate that the particles are stable, red squares unstable. Adapted with permission from Ref.[75] (Copyright 2014 American Chemical Society).*

Here we present stability results of 6.8 and 13.2 nm γ-$Fe_2O_3$ NPs coated with $(MPEG_{2k})_3$-*co*-$MPh_3$ and $(MPEG_{5k})_7$-*co*-$MPh_7$ in various solvents.[75,78] **Fig. 5a** displays the hydrodynamic diameter $D_H$ over 7 days after their dilution in DMEM complemented with serum. Upon NP injection, $D_H$ jumps to its value found in DI-water and remains stationary over the entire period. Subsequent measurements at 1 month confirm the long-term stability of the PEGylated NPs. In contrast, experiments performed on bare or citrate coated NPs, as well as on commercial MRI contrast agent Cliavist® exhibit a continuous increase after injection, indicating aggregation (**Fig. 5b**). The mechanisms responsible for aggregation are electrostatic screening due to physiological conditions and ligand exchange.[90] The stability results obtained with $\gamma$-$Fe_2O_3$ coated with various copolymers, including $(MPEG_{2k})_4$-*co*-$MA_4$ where functional groups were methacrylic acids are summarized in **Fig. 5c** for PBS1X, 1 M NaCl DI-water, DMEM and DMEM with serum. Following these first measures, similar studies were conducted on $CeO_2$-NPs, $TiO_2$ nanoclusters and $Al_2O_3$ platelets with sizes between 7 to 40 nm, and confirmed the merits of $MPEG_{ik}$-*co*-MPh copolymers for coating. In this study, $MPEG_{ik}$-*co*-$MPEG_{jk}$a-*co*-MPh terpolymers (**Fig. 2d**) allowing subsequent functionalization *via* an amine coupling were evaluated and found to be efficient as well for colloidal stability. Interestingly, it was found that the PEG layer thickness for a given PEG length was identical whatever the oxide core (i.e. made from iron, cerium, titanium and aluminum) and that NPs exposed to serum proteins were devoid of corona.[46] With $MPEG_{ik}$-*co*-MPh polymers, optimum coatings were with PEG densities of 1 $nm^{-2}$ and layer thickness about 10 nm.[14,30,91]

## ■ POLYMERS ON FLAT SURFACES, BRUSH FORMATION AND PROTEIN RESISTANCE STUDIED BY QCM-D

### Phosphonic acid PEG copolymer adsorption on $Fe_3O_4$ substrate

In the previous sections, we outlined protocols in which $MPEG_{ik}$-*co*-MPh copolymers spontaneously adsorb onto MOx-NPs surfaces, leading to the formation of a resilient protective coating. Here we pursue this approach by extending it to flat surfaces using quartz crystal microbalance with dissipation (QCM-D), a technique that allows to monitor the kinetics of adsorption with high sensititvity.[92] A QCM-D equipment was used in conjunction with quartz crystal sensors coated by a thin iron oxide ($Fe_3O_4$) film. Polymer solutions (1 g $L^{-1}$, pH 2.0) were injected in an exchange mode into the chamber containing the crystal sensor. Any substance adsorbing onto the crystal surface induces a decrease of the resonance frequency $f_n = f_0 + \Delta f_n$ ($\Delta f_n$ < 0) with respect to the fundamental resonance frequency $f_0$, $n$ being the $n^{th}$-overtone. $\Delta f_n$ is related to the adsorbed mass per unit area (ng $cm^{-2}$) through the Sauerbrey equation, $\Delta m = -C\,\Delta f_n/n$ where $C$ = 17.7 ng s $cm^{-2}$ is the Sauerbrey constant.[93] Of note, the adsorbed mass also includes the water trapped hydrodynamically inside the layer. **Fig. 6a** displays the real-time kinetics of polymer adsorption for $(MPEG_{1k})_4$-*co*-$MPh_4$, $(MPEG_{2k})_3$-*co*-$MPh_3$ and $(MPEG_{5k})_7$-*co*-$MPh_7$ at pH 2.0. After injection, the areal mass density $\Delta m$ increases rapidly and reaches a stationary value that depends on the PEG molecular weight. The QCM-D data show no change in mass density upon rinsing with DI-water at pH 7.4, indicating that the polymers are strongly bound to the $Fe_3O_4$ surface. With increasing molecular weight, $\Delta m$ increases from 400 to 750 ng $cm^{-2}$, which is a typical feature for polymers adsorbing at interfaces *via* grafting-to mechanisms.[91,94-96] To retrieve the thickness $h_{2D}$ of the adsorbed layer, the Voigt model[97] was applied to the dissipation data, leading to values of 4.1, 5.7 to 9.4 nm for $(MPEG_{1k})_4$-*co*-$MPh_4$, $(MPEG_{2k})_3$-*co*-$MPh_3$ and $(MPEG_{5k})_7$-*co*-$MPh_7$ respectively.[62] These values are in agreement with those of adsorbed polymers of similar molecular weights,[91,96] and compare well with the coating thickness $h_{NP}$ obtained on NPs.[75,78]

To estimate the PEG density $\sigma_{PEG}$, the polymer brush theory was applied.[28-29,98] In the moderate and high surface density regimes, $\sigma_{PEG}$ and $h_{2D}$ are linked through the relation:[30,99]

$$h_{2D} = \left(\frac{\sigma_{PEG}}{3}\right)^{1/3} b^{2/3} a N \qquad (2)$$

where $a$ = 0.28 nm and $b$ = 0.72 nm are respectively the chemical monomer and Kuhn lengths for PEG[91] and $N$ the degree of polymerization of the chains. Eq. 2 leads to PEG densities of 1.55, 0.57 and 0.15 $nm^{-2}$ for $(MPEG_{1k})_4$-*co*-$MPh_4$, $(MPEG_{2k})_3$-*co*-$MPh_3$ and $(MPEG_{5k})_7$-*co*-$MPh_7$. A commonly used parameter for quantitative characterization of polymer brushes is the reduced tethered density $\Sigma = \pi\sigma_{PEG}R_g^2$.[100] From the QCM-D data, $\Sigma$ is found in the range 3.8 to 11.7, corresponding to grafting densities in the moderate (methacrylic acid-$PEG_{2k}$, $1 < \Sigma < 5$) and in the highly stretched regimes (phosphonic acid-$PEG_{1k}$, $PEG_{2k}$ and $PEG_{5k}$, $\Sigma \geq 5$).[30,101-102] A schematic representation of a stretched PEGylated layer is illustrated in **Fig. 6b**. A 10-fold decrease in $\sigma_{PEG}$ between $(MPEG_{1k})_4$-*co*-$MPh_4$ and $(MPEG_{5k})_7$-*co*-$MPh_7$ is attributed to excluded volume interaction and steric repulsion between chains during deposition. The already adsorbed chains act as a barrier to the incoming ones, a mechanism that is more effective for longer chains. As a result, the brush stretching and morphology are different: dense and solid-like for $PEG_{1k}$ and soft and viscoelastic for $PEG_{5k}$.[91]

a
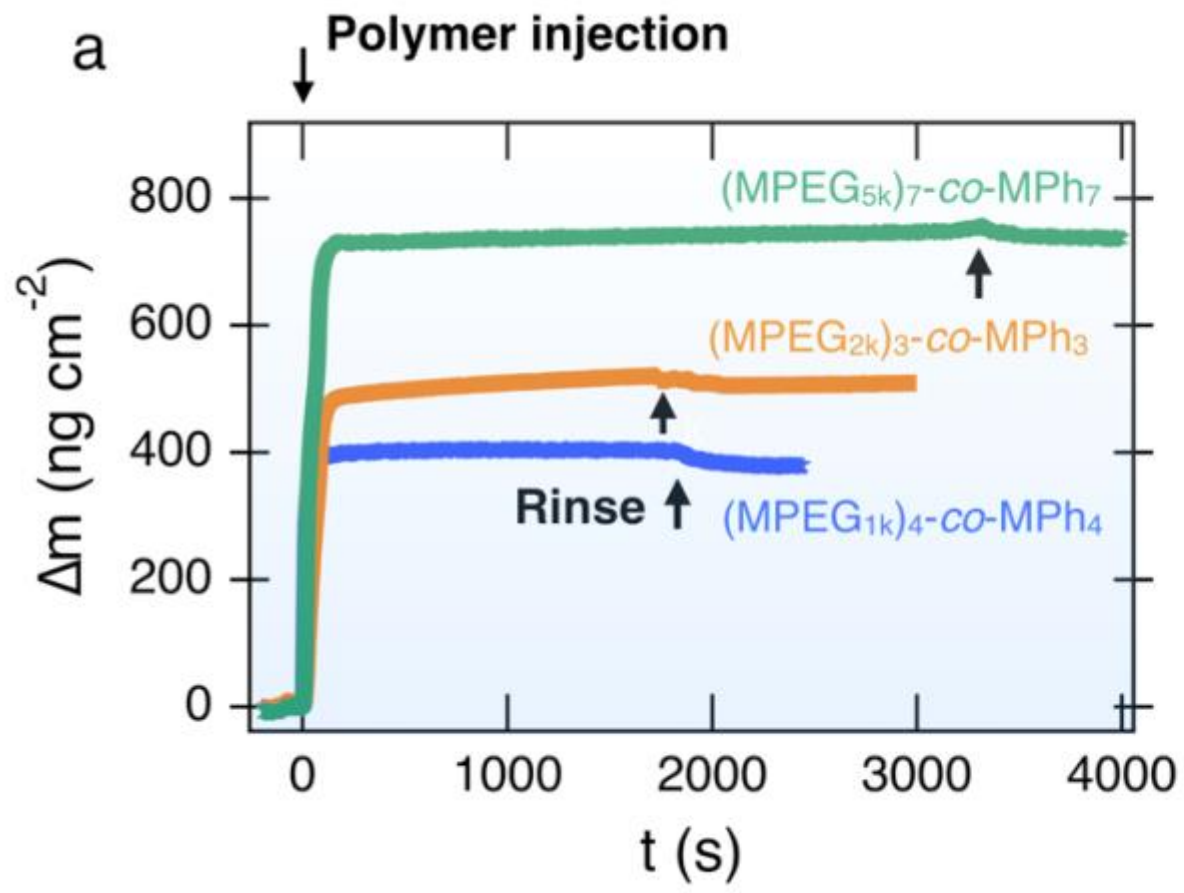

b
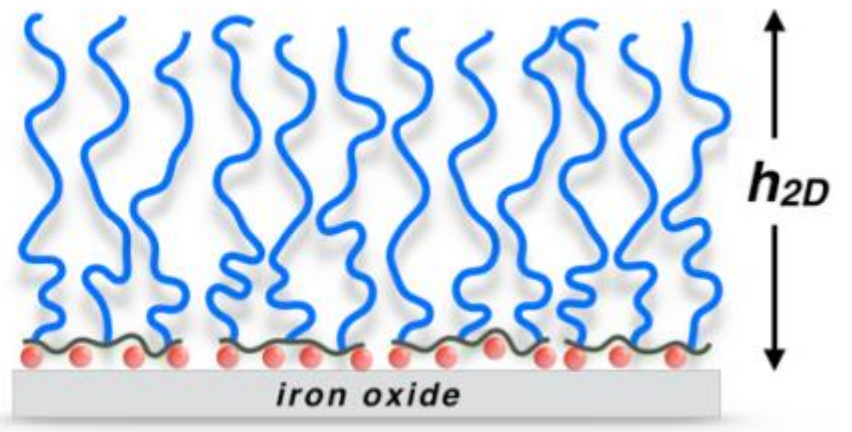

c

| Coating | $h_{2D}$ (nm) | $\sigma_{PEG}$ (nm$^{-2}$) |
|---|---|---|
| $(MPEG_{1k})_4$-*co*-$MPh_4$ | 4.1 | 1.55 |
| $(MPEG_{2k})_3$-*co*-$MPh_3$ | 5.7 | 0.53 |
| $(MPEG_{5k})_7$-*co*-$MPh_7$ | 9.4 | 0.15 |
| $(MPEG_{2k})_4$-*co*-$MA_4$ | 4.6 | 0.28 |

***Figure 6**: **a)** Real-time binding curves for the areal mass density $\Delta m(t)$ obtained from QCM-D experiments during the adsorption of the copolymers $(MPEG_{1k})_4$-co-$MPh_4$, $(MPEG_{2k})_3$-co-$MPh_3$ and $(MPEG_{5k})_7$-co-$MPh_7$ on $Fe_3O_4$ substrates at pH 2.0. Experimental conditions were T = 25 °C and copolymer concentrations of 1 g $L^{-1}$. The crystal was excited at the resonance frequency $f_0$= 4.95 MHz through a voltage applied across the gold electrodes. $\Delta m$ was derived from the third overtone of the resonance frequency $\Delta f_3/3$ using the Sauerbrey equation. Deposition at pH 7.4 led to mitigated results and unstable polymer layers.[62] **b)** Schematic representation of a stretched PEGylated layer deposited on iron oxide substrates. **c)** Summary of phosphonic acid PEG copolymers assessed with QCM-D, including the layer thickness $h_{2D}$ and PEG density $\sigma_{PEG}$.*

**Resistance of PEGylated layers to protein adsorption**

Once the polymers firmly adsorbed, we evaluated the resistance of the coating layers to proteins by injecting a physiological solution (PBS) containing fetal bovine serum (FBS) at 10 vol. % to the QCM-D chamber. **Fig. 7a** displays the real-time kinetics of protein adsorption illustrated by the areal mass densities $\Delta m(t)$ for three substrates: *i)* uncoated iron oxide, *ii)* iron oxide coated with a high density 4.1 nm $(MPEG_{1k})_4$-*co*-$MPh_4$ layer, and *iii)* iron oxide coated with the high density 9.4 nm $(MPEG_{5k})_7$-*co*-$MPh_7$ layer (**Fig. 7b**). After FBS injection on the uncoated substrate, $\Delta m(t)$ increases rapidly and levels off at 1020 ng $cm^{-2}$ at steady state, with no modification upon rinsing with DI-water. This outcome indicates that proteins are strongly bound to the $Fe_3O_4$ film, confirming the strong affinity of proteins for untreated iron oxide surfaces.[91,95,103]

In the presence of PEG brushes, the protein adsorption behavior changes drastically. Stationary areal mass densities $\Delta m$ for $(MPEG_{1k})_4$-*co*-$MPh_4$ and $(MPEG_{5k})_7$-*co*-$MPh_7$ are lower than those obtained with the uncoated surface. After rinsing, the residual mass densities is decreased 3 fold for $(MPEG_{1k})_4$-*co*-$MPh_4$, whereas it is reduced 100-fold for $(MPEG_{5k})_7$-*co*-$MPh_7$, to a value that is close to the QCM-D detection limitc (~ 2 ng cm$^{-2}$). For the latter sample, QCM-D data suggest that the proteins are weakly attached to the PEG layer and removed by rinsing. Another crucial result is that the FBS addition did not modify the structure of the PEGylated layer, as fequency and dissipation return to their pre-FBS injection levels.

**Fig. 7c** displays an histogram of the protein areal mass densities deposited on PEG brushes before and after rinsing. The data also include the stationary $\Delta m$ for the methacrylic acid analogue, $(MPEG_{2k})_4$-*co*-$MA_4$. While $(MPEG_{2k})_3$-*co*-$MPh_3$ behaves like its 5 kDa counterpart, methacrylic acid containing polymers are less efficient against protein adsorption. Upon FBS injection, data show a continuous and irreversible degradation of the PEG layer due the competitive complexation between proteins and grafted copolymers. The protein adsorption resistance coefficient for $MPEG_{ik}$-*co*-MPh are 82%, 98% and 99% for i = 1, 2 and 5 respectively, whereas it is only 65% for $(MPEG_{2k})_4$-*co*-$MA_4$, demonstrating that best anti-fouling layers are made from copolymers with multiple phosphonic acid functional groups.

In conclusion of this section, we have established a parallel in the deposition of $MPEG_{ik}$-*co*-MPh copolymers on highly curved interfaces and on flat QCM-D substrates. In both, the copolymers spontaneously adsorb to the iron oxide interfaces at acidic pH and form a resilient nm-thick PEG brush. The copolymer backbone attaches to the surface *via* multisite bonds, and the PEG side chains organize into stretched protein-resistant brushes.

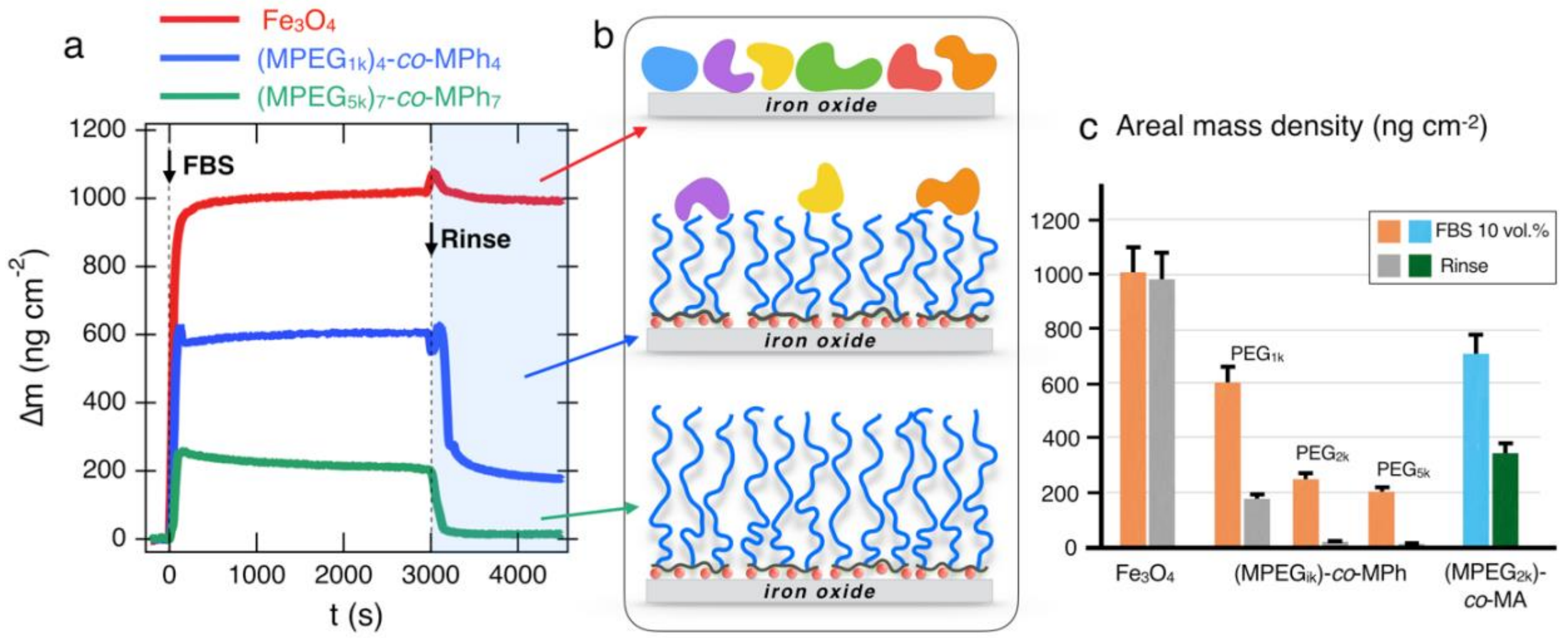

***Figure 7: a)*** *QCM-D responses of uncoated $Fe_3O_4$, $Fe_3O_4$@$(MPEG_{1k})_4$-co-$MPh_4$ and $Fe_3O_4$@$(MPEG_{5k})_7$-co-$MPh_7$ substrates exposed to a 10 vol. % fetal bovine serum (FBS) solution. FBS contains mostly bovine serum albumin (BSA) and globulin proteins at a total concentration of 40 g L$^{-1}$. The areal mass density $\Delta m$ is obtained from the Sauerbrey equation. Arrows indicate FBS injection and rinsing.* ***b)*** *Illustration of the different adsorption and rinsing stages used in QCM-D, including exposition to serum proteins and rinsing for uncoated $Fe_3O_4$ (upper panel), $Fe_3O_4$@$(MPEG_{1k})_4$-co-$MPh_4$ (middle panel) and $Fe_3O_4$@$(MPEG_{5k})_7$-co-$MPh_7$ (lower panel) substrates.* ***c)*** *Protein areal mass densities $\Delta m$ on coated and uncoated $Fe_3O_4$ substrates.*

## ■ APPLICATIONS TO NANOMEDICINE AND BIOMEDICAL IMAGING

### Effect of coating on In Vitro cellular uptake

Issues dealing with the impacts, fate and toxicity of NPs have become the focus of many recent researches.[104-105] Despite these efforts, the risk assessment of nanomaterials, including MOx-NPs, to human health and the environment has not been fully evaluated. The objectives of toxicological assays are the measure of the viability of living cells incubated with NPs, and of the half-inhibitory concentration, which quantifies the dose to inhibit cell growth by half after 24 hours. These studies have shown that NP cytotoxicity depends on several factors, two of which play a determining role: the amount or mass of internalized NPs at a given dose, and their localization in the cytoplasm. Regarding the uptaken amount for MOx-NPs, metal detection techniques based on mass or light spectrometry now allow an accurate measure of internalized material per cell, with an accuracy better than $10^3$ to $10^4$ NP units, or ~ 10 femtograms in mass. These mass-per-cell and localization results provide valuable information about the mechanisms of cell-NPs interaction,[106-107] which can then be used to design In Vivo strategies regarding organ or tissue targeting.

We illustrate here a specific feature of the NP-cell interaction, namely the role of a PEGylated coating in the stealthiness of NPs towards cells. $CeO_2$ and $\gamma$-$Fe_2O_3$-NPs presented in the previous sections are again objects of interest and we focus on various coatings, including citrate ligands, poly(acrylic acid) and $MPEG_{ik}$-*co*-MPh copolymers. In these In Vitro assays, the tested cells are of murine origin and included bEnd.3 brain epitethlial cells, 3T3/NIH fibroblasts and RAW 264.7 macrophages. For $CeO_2$-NPs, the inductively coupled plasma - optical emission spectrometry (ICP-OES) technique was used, whereas for $\gamma$-$Fe_2O_3$-NPs we harnessed the MILC (for Mass of metal Internalized/Adsorbed by Living Cells) protocol, which is based on the digestion of cells by hydrochloric acid and colorimetric determination.[75,108] **Fig. 8a** displays the mass of cerium $M_{Ce}$ uptaken by bEnd.3 cells expressed in picogram per cell as a function of the $CeO_2$ dose in the cell culture medium. It is found that both coated and uncoated $CeO_2$-NPs associate with bEnd.3 cells after a 24 h incubation. However, the amount of $CeO_2$@$(MPEG_{2k})_5$-*co*-$MPh_5$ detected is significantly reduced, by about 100-fold, compared to bare $CeO_2$. The straight line in dark grey in the figure represents the maximum amount of cerium that can be uptaken by a single cell under current exposure conditions. It is seen that the percentage decreases from 20-30% for the bare NPs to 1‰ for the coated ones. In these assays, it was shown that the few PEGylated NPs that crossed the cell membrane were found in endosomes, suggesting an entry mechanism *via* endocytosis.[107] Similar results were obtained by coating $CeO_2$-NPs with $MPEG_{ik}$-*co*-$MPEG_{jk}$a-*co*-MPh terpolymers (Fig. 2d) containing an amine-terminated PEG allowing subsequent chemcial conjugation.[109]

**Fig. 8b** and **Fig. 8c** display similar types of results for the mass of iron $M_{Fe}$ uptaken by NIH/3T3 fibroblasts and RAW 264.7 macrophages, respectively. The cells were treated for 24 h with 13.2 nm $\gamma$-$Fe_2O_3$ NPs coated with citrate, $PAA_{5k}$ and $(MPEG_{2k})_3$-*co*-$MPh_3$ at concentrations between $10^{-3}$ and 1 g $L^{-1}$. For citrate coated particles, the masses of internalized/adsorbed iron were high (about 50-70 pg/cell) and about 10-20% of the maximum value (straight dark line). For particles coated with $PAA_{5k}$, the behavior is similar, however with a lower saturation plateau. Conversely, with $(MPEG_{2k})_3$-*co*-$MPh_3$ copolymers as a coat, the uptake levels are of the order of 0.1 pg/cell, *i.e.* 50 to 100 times lower those found with citrate and poly(acrylic acid). The results in **Fig. 8** highlight the stealthiness of PEGylated NPs, regardless of their chemical nature or the cell line studied. To understand this phenomenon, it is important to remember uncoated NPs or NPs coated with citrate ligands aggregate in culture media (**Fig. 5**). NP aggregates are prone to adsorb at the cellular membrane by sedimentation, resulting in a concentration gradient above the cell layer. Sedimentation-dosimetry models[110] have showed that for micron-sized aggregates, the effective dose seen by the cells can exceed the nominal one by a factor of 10 to 1000, thereby increasing cell death and toxicity.[111] These outcomes evidence the crucial impact of the coating in the evaluation of NP-cells interaction and cytotoxicity.

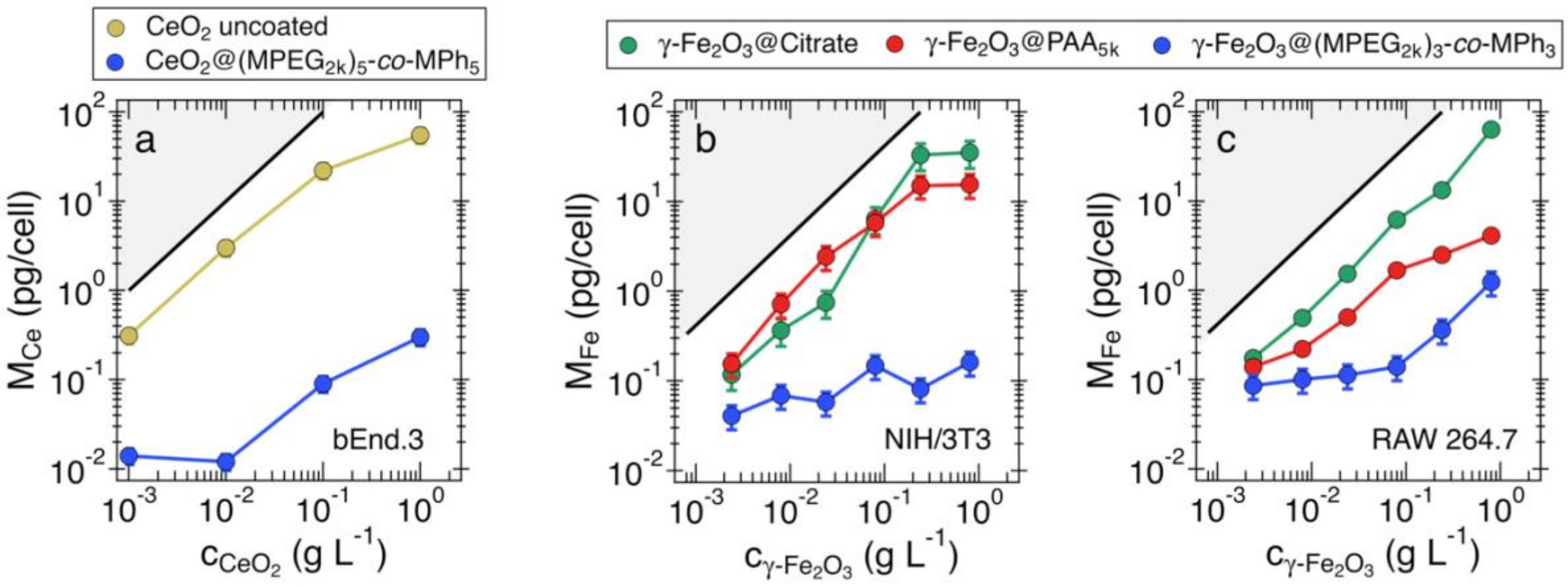

***Figure 8:*** *Mass of metal atom $M_{Ce}$ and $M_{Fe}$ per cell expressed in pg/cell as a function of the metal oxide concentration in the cell culture medium after a 24 h incubation:* ***a)*** *cerium oxide;* ***b,c)*** *iron oxide. The eukaryotic cell lines studied are from murine origin and include bEnd.3 brain epitethlial cells, 3T3/NIH fibroblasts and RAW 264.7 macrophages. The $CeO_2$ and $\gamma$-$Fe_2O_3$ NPs have diameters of 7.8 nm and 13.2 nm and are those already discussed in* ***Section II.1****. For these particles, 1 pg of metal corresponds to 2.5×10$^6$ and 5×10$^4$ NP units, respectively. As coatings, we used citrate ions, poly(acylic acid) of $M_w$ = 5 kDa, and two phosphonic acid PEG copolymers, $(MPEG_{2k})_3$-co-$MPh_3$ of $M_w$ = 13 kDa[75] and $(MPEG_{2k})_5$-co-$MPh_5$ of $M_w$ = 20 kDa[109] The straight thick lines in each figure depict the maximum amount of MOx-NPs taken up by a single cell. Adapted with permission from Ref.[109] (Copyright 2021 Wiley-VCH GmbH) and Ref.[75] (Copyright 2014 American Chemical Society).*

**Effect of coating on In Vivo nanoparticle pharmacokinetics monitored by time-resolved MRI**

In medicine, the pharmacokinetics consists in determining the fate of substances administered to a living organism intravenously, and its kinetics of accumulation and clearance in various organs, such as the liver, spleen, lungs, kidney. Studies have shown that NPs administered in the blood compartment are opsonized by plasma proteins, preventing them to interact specifically with potential targets.[9,112] At the same time, protein adsorption activates the NP uptake by the mononuclear phagocytic system *via* circulating macrophages and monocytes. The two-step opsonization mechanism is responsible for the NP clearance from the bloodstream and for their accumulation in unrelated organs, typically the liver and the spleen.[86] Following the In Vitro assays, we undertook a series of pharmacokinetics studies of coated $\gamma$-$Fe_2O_3$-NPs using an MRI 7-Tesla spectrometer. $\gamma$-$Fe_2O_3$@coating dispersions were injected intravenously into the tail vein of wild-type female 8 weeks BALB/c mice at a dose of 1 mg of iron per kilogram of mouse. Spatially resolved images of the liver, spleen and kidneys were acquired prior and after injection, at time points between one minute and seven days. The change in the liver MRI contrast was investigated as it provided direct insight into the NP pharmacokinetics. With this protocol, the time course of NP uptake and clearance in and out of the liver could be retrieved.

**Fig. 9a** and **9b** compares MRI scans obtained in the first three hours for the benchmark Cliavist® and for $\gamma$-$Fe_2O_3$@$(MPEG_{2k})_3$-*co*-$MPh_3$. Commercialized by Bayer for liver imaging, Cliavist® contains partialy aggregated 4 nm magnetic core particles coated with 1.4 kDa carboxydextran polymers.[82] With this benchmark, the mouse liver section exhibited a negative contrast enhancement (darkening) already 5 minutes post i.v., showing that NPs accumulated in the liver.[81-82] For $\gamma$-$Fe_2O_3$@$(MPEG_{2k})_3$-*co*-$MPh_3$, on the contrary the darkening was delayed by about 3 hours. During the period, the MR images showed no contrast change of the liver and of other organs (spleen, kidneys), an indication of a prolonged circulation in the bloodstream. At 3 hours post i.v., MRI images exhibited a negative contrast enhancement, although less intense than with Cliavist®. To quantify the MRI images, the liver mapping was performed by integrating the grey intensity $I_{MRI}(t)$ on regions of interest (continuous lines in **Figs. 9a** and **9b**). Spatially average intensities were normalized with respect to the pre-injection level $I^0_{MRI}$ and translated the quantity $1 - I_{MRI}(t)/I^0_{MRI}$. In Ex Vivo calibration assays, $1 - I_{MRI}(t)/I^0_{MRI}$ was found to vary linearly with the $\gamma$-$Fe_2O_3$-NPs concentratrion and we hypthesize that this behavior holds also In Vivo (although the iron concentrations for each assay are obtained in relative units). The results of this procedure averaged on $n$ mice are shown for cliavist® ($n$ = 3), $\gamma$-$Fe_2O_3$@$PAA_{2k}$ ($n$ = 3) and $\gamma$-$Fe_2O_3$@$(MPEG_{2k})_3$-*co*-$MPh_3$ ($n$ = 4) in **Figs. 9c-e** respectively. For Cliavist® and $\gamma$-$Fe_2O_3$@$PAA_{2k}$, the iron oxide concentration in the liver increased rapidly after injection and exhibits a broad maximum around 200 minutes. At longer timescale, from D1 to D7 days post i.v., the concentration fell to 0, indicating the complete iron clearance from the liver.[113-114] On the opposite, $\gamma$-$Fe_2O_3$@$(MPEG_{2k})_3$-*co*-$MPh_3$ exhibited prolonged circulations in the blood, the concentration increase appearing approximately 2-3 hours post i.v.. The concentration maximum was shifted to 1000 minutes, after which iron was cleared from the liver in a few days, as for the previous NPs.

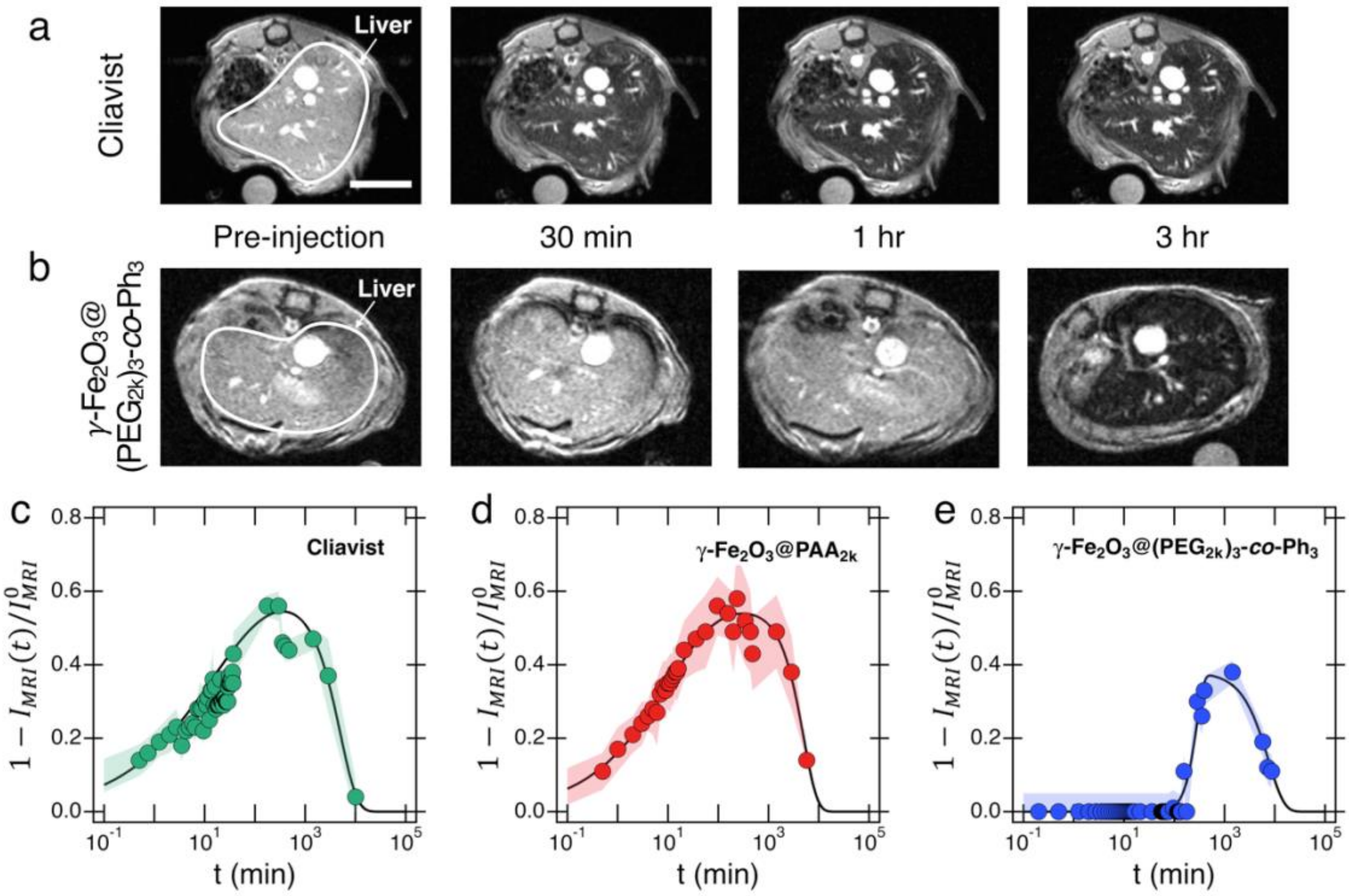

***Figure 9:*** *Magnetic resonance imaging scans for wild-type female 8 weeks BALB/c mouse livers at different time points, 30 min, 1 and 3 hours after injection for (**a**) Cliavist® and for (**b**) 6.8 nm iron oxide particles coated with $(MPEG_{2k})_3$-co-$MPh_3$. The scale bar in **a** is 5 mm. **c-e)** Time dependence of the quantity $1 - I_{MRI}(t)/I^0_{MRI}$ for Cliavist®, $\gamma$-$Fe_2O_3$@$PAA_{2k}$ and $\gamma$-$Fe_2O_3$@$(MPEG_{2k})_3$-co-$MPh_3$ respectively where $I_{MRI}(t)$ is the MRI contrast and $I^0_{MRI}$ its preinjection value. From calibration experiments in solution, it was shown, that within a multiplicative factor $1 - I_{MRI}(t)/I^0_{MRI}$ is proportional to the iron concentration in the liver (within a multiplicative factor). Injected dose was 1 mg of iron per kilogram of mouse.*

To retrieve the pharmacokinetics parameters, it was assumed that the NPs capture and the clearance mechanisms were occurring sequentially, and that the concentration was the addition of two stretched exponentials characterized by half-life times $t_{1/2}^{up}$ and $t_{1/2}^{clear}$, respectively. As a result, the uptake time was estimated at 6.0 and 4.5 min for Cliavist® and $\gamma$-$Fe_2O_3$@$PAA_{2k}$, whereas it was multiplied by about 50 for $\gamma$-$Fe_2O_3$@$(MPEG_{2k})_3$-*co*-$MPh_3$ at $t_{1/2}^{up}$ = 250 min, confirming the strong dependence on the coating. As for the clearance time, it was found that $t_{1/2}^{clear}$ did not depend on the magnetic core sizes, nor on the coating, with half-life times between 2 to 4 days. Regarding this last mechanism, it was assumed that the clearance occurred *via* the particle degradation and iron incorporation into the main intracellular storage proteins, e.g. ferritin and hemosiderin.[113-114] The main outcome from the In Vivo MRI mouse assays is that coating appears as an important parameter that affects the liver uptake kinetics from the blood pool. $MPEG_{ik}$-*co*-MPh copolymers provides the most efficient contrast agent in terms of stealthiness, as they are able to prolong the blood particle lifetime for several hours. This result could be exploited for passive targeting of cancer drugs using the mechanism of enhanced permeability and retention effect (EPR).

## ■ THERMOSENSITIVE IRON OXIDE NANOPARTICLES

So far, we have focused on a biomedical theme, but the above strategies can easily be combined with a material science approaches, and the development of stimuli-responsive or smart materials. Smart materials refer to compounds that respond to an external stimulus or to the environment with a dynamic and controlled change on their properties. The classical stimuli harnessed in physicochemistry are pH, ionic strength, chemical composition, temperature, light, magnetic and electric fields.[18,115-116] Herein we exploited our coating platform and apply it to 9.4 nm $\gamma$-$Fe_2O_3$-NPs to design novel core-shel particles that are temperature and magnetic field stimuli responsive.[117-118] For that, we used the 49.5 kDa copolymer $(MPAm)_{424}$-*co*-$MPh_9$ in which the PEGylated chains have been substituted by thermosensitive propylacrylamide groups (PAm) (**Fig. 10a**). Water soluble thermoresponsive polymers such as poly(N-isopropylacrylamide) (PNIPAm) or its isomer poly(N-n-propylacrylamide) (PNnPAm) have a low critical solution temperature (LCST). With increasing temperature above the LCST, these compounds undergo a reversible phase transition from a soluble hydrated state to an insoluble dehydrated state.[119] PNnPAm has an LCST around 23 °C, which is characterized by a steeper transition and more pronounced heating/cooling hysteresis compared to PNIPAm.[120-121]

The $\gamma$-$Fe_2O_3$-NP coating protocol was here identical to that of the previous sections, the only difference being that mixing was carried out at 5 °C, a temperature where the $(MPAm)_{424}$-*co*-$MPh_9$ chains are well solvated. When pH was brought to pH 8.0, a critical mixing ratio at $X_C$ separating stable dispersions and precipitates in the phase behavior was disclosed. The $X_C$ value of 0.5 leads to an average with 54 copolymers or equivalently 430 phosphonic acid moities per particle (**Fig. 10b**). TEM and light scattering experiments shows that following the coating process the NPs were slightly agglomerated (**Fig. 10c**), with hydrodynamic sizes in the order of 80 nm, as compared to 25 nm for the bare NPs. This partial aggregation has been explained by the fact that phosphonic acids from the same chain were able to graft different $\gamma$-$Fe_2O_3$-NPs, and bridged them together. Electrophoretic mobility measurements also provided negative zeta potential of the order of -25 ± 5 mV, which was attributed to ungrafted phosphonic acid groups at the surface (**Fig. 10b**).

To study the thermoresponsive behavior of γ-$Fe_2O_3$@$(MPAm)_{424}$-*co*-$MPh_9$, the dispersions were salted to screen the electrostatic interaction arising from non-adsorbed $PO_3^{2-}$ groups. At pH 8.0 and [NaCl] = 200 mM, the LCST was found to be shifted by 5 °C compared to single polymers, at 28 °C.[122] **Fig. 10d** displays a representative TEM image of the NPs above the LCST, here at 40 °C, where a noticeable aggregation can be observed. A closer look at this image allows to identify patches around the NPs corresponding to collapsed polymers. **Fig. 10e** shows the time dependence of the hydrodynamic diameter upon heating from 5 to 40 °C. $D_H(t)$ increases linearly and then saturates around 200 nm in the characteristic time of 40 s, confirming that γ-$Fe_2O_3$@$(MPAm)_{424}$-*co*-$MPh_9$ self-assembled well under temperature. Upon cooling down to 10 °C, the initially formed aggregates disassembled. However, the disassembly process occurs on a time scale of a few hours, the complete process taking a few days. In this regime, $D_H(t)$ follows a power law of the form $D_H(t) \sim t^{0.06}$ (**Fig. 10f**). Hystereses in heating and cooling cycles have been reported in the literature for thermoresponsive NPs[122] and are explained in terms of differences in transition enthalpies. It was also suggested that the diffusion of water into the aggregated polymer network is slower due to the steric hindrance imparted by the entangled chains.[122] An asymetric hysteresis in heating/cooling cycles are considered as a drawback for applications.[123] Here we show that it can represent a definite advantage for the design of nano- or micro-functional objects. **Fig. 10g** and **10h** show optical microscopy images of magnetic filaments resulting from the combined application of temperature ($T > LCST$) and magnetic field (B = 300 mT) to a γ-$Fe_2O_3$@$(MPAm)_{424}$-*co*-$MPh_9$ dispersion.

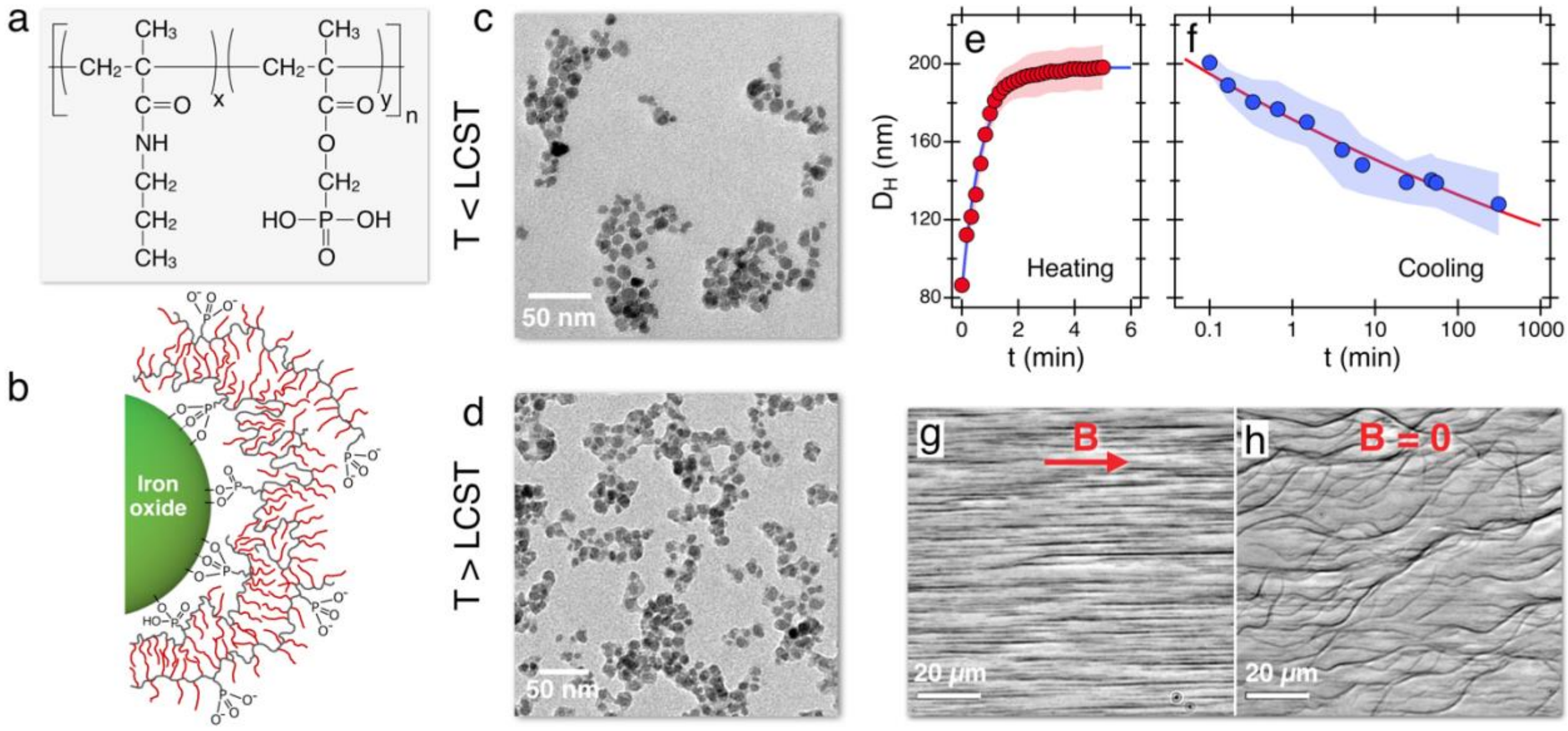

***Figure 10**: **a)** Chemical formula of the thermosensitive copolymer $(MPAm)_{424}$-co-$MPh_9$ of weight-averaged molecular weight 49.5 kDa. The comonomer proportions are $x$ = 0.98 and $y$ = 0.02, for a degree of polymerization of $n$ = 432 . **b)** Schematic representation of the polymer coating layer around a 9.4 nm iron oxide nanoparticle. **c)** and **d)** Transmission electron microscopy images of γ-$Fe_2O_3$@$(MPAm)_{424}$-co-$MPh_9$ dispersions below and above the LCST. **e)** and **f)** Time dependence of the hydrodynamic diameter $D_H(t)$ in the temperature-driven assembly from and disassembly processes, respectively. Temperature jumps are from 5 °C to 40 °C (heating), and from 40 °C to 10 °C (cooling). **g)** and **h)** Phase-contrast optical microscopy images of magnetic filaments under the application of the magnetic field, and after the removal of the field. Adapted with permission from Ref.[118] (Copyright 2021 Springer Nature).*

When magnetic NPs are submitted to a magnetic field, strong dipole-dipole magnetic interactions induce the formation of chains.[124] The simultaneous heating promotes the phase transition of the polymer shells, leading to a collapse of the hybrid structure and the formation of magnetic filament. It is found that these filaments retain their microstructure after switching off the field or cooling down the sample to room temperature (**Fig. 10h**). The resulting magnetic filaments can be washed, isolated, and used for further experiments. They also exhibit a response to an applied magnetic field and can rotate, bend, or move under the application of an external magnetic field.[117-118] In conclusion, we have shown here a novel physicochemical method for the preparation of bio-inspired materials using a template-free controlled assembly. Taking advantage of the hysteresis of the cooling-heating transition of PNnPAm coating, we fabricated long-lived one-dimensional microstructures in the form of magnetic microfilaments that could find potential applications in remotely controlled devices, such as artificial cilia for microfluidic pumping.[125-126]

## ■ PERSPECTIVES AND FUTURE DIRECTIONS/CONCLUSIONS

In this review, we provide a summary of our research on phosphonic acid PEG copolymers as coating material for metal oxide NPs. Surface modification is a rapidly growing field that is still being extensively studied today. The approach presented here is at the crossroads of self-assembled monolayers and the Layer-by-Layer assembly, which are considered as the most advanced techniques for non-covalent coating.[3] It makes use of phosphonic acid motifs, which are common functional groups in SAMs and it takes advantage of the electrostatic complexation between the opposite charges carried by NPs and copolymers. In contrast to L-*b*-L approach however, there is only one layer of polymers deposited and its thickness depends on the degree of polymerization of the water-soluble chains, and not on the number of deposited layers. From the recent literature on MOx-NP coating, it can be seen that with a few exceptions[69] functional polymers engineered for surface modification are designed for a single type of substrate and for a predetermined application. Nowadays, there is a need for coatings that can be applied for a wide range of surfaces, including those of NP dipersions. Here, we highlight that the present coating platform can fill this gap and contribute to the development of versatile multi-substrate coatings.

This platform is based on a series of elementary schemes that could rationally be reproduced with polymers and NPs of different chemistries. The first scheme is the synthesis by free radical polymerization of MPEG$_{ik}$-*co*-MPh statistical copolymers that contain multiple phosphonic acid functional groups. In addition to electrostatic complexation that drives the initial adsorption,[62] as shown by QCM-D (**Fig. 6**), the presence of three oxygen atoms on the phosphonic acid allows for mono-, bi- and tri-dentate binding coordination modes. The second basic principle is the use of the *two-step coating process* where NPs and polymers are synthesized separately, and later assembled following relevant protocols. In the survey, we placed some emphasis on the assembly protocols and displayed several NP@coating behaviors, such as those depicted in **Figs. 3** and **4**. In the course of our studies, we discovered that these protocols were versatile enough to be transposed to a wide variety of particles and polymers. As shown by light scattering (**Fig. 5**) and QCM-D (**Fig. 6**), MPEG$_{ik}$-*co*-MPh copolymers spontaneously adsorb onto the MOx surfaces investigated ($CeO_2$, $\gamma$-$Fe_2O_3$, $Fe_3O_4$, $TiO_2$ and $Al_2O_3$), leading to the formation of a nanometer thick and dense PEGylated brush. The designed platform allows moreover to evaluate the intrinsic performance of an adlayer. On the one hand, we can start from a given NP, and build around it multiple coatings to disclose which of them is the most appropriate for a given application. We did this for iron oxide contrast agents In Vivo and based on pharmacokinetic studies, it is found that $PEG_{2k}$ layers are the most effective in terms of prolonged circulation in the blood compartment of mice.[45,78] Interestingly, the QCM-D data further confirmed that $PEG_{2k}$ and $PEG_{5k}$ coatings were resistant to serum proteins. This approach also demonstrated that polymers with a unique phosphonic acid functional group are poor colloidal stabilizers compared to copolymers with multiple anchors.[46] We can also reason in the reverse way: starting from a given polymer, we can screen a large library of NPs to set up the fundamental rules of adsorption on nanometric surfaces. This was systematically done with $CeO_2$, $\gamma$-$Fe_2O_3$, $TiO_2$ and $Al_2O_3$ NPs where we could show that on NPs of various morphologies, the adlayer had kept the same thickness, pointing to a common deposition mechanism.[46] Pertaining to copolymers in which the PEGylated chains have been substituted by thermosensitive propylacrylamide groups, we have been able to modulate the assembly properties of iron oxide particles, and form flexible filaments.[117-118] In the perspective of applying phosphonic acid-based polymers to a wider range of surfaces, it is worth mentioning recent attempts to stabilize other core materials, such as luminescent quantum dots and gold particles.[127-128]

As for protein adsorption on NPs or flat substrates, we confirmed a theroretical result demonstrated 30 years ago, namely that PEGylated brushes with density around 1 $nm^{-2}$ and thickness 10 nm are anti-fouling and protein resistant. Our results also suggest that protein corona is not a ubiquitous phenomenon, as commonly believed, since all four coated metal oxide NPs studied remain stable in protein-rich media, and are devoid of proteins at their surface over long period of time (> months). We also established a parallel in the deposition of $MPEG_{ik}$-*co*-MPh copolymers between highly curved interfaces and a flat QCM-D substrate, an outcome that is interesting for assessing new functional polymers.

Regarding the development of novel polymers, the polymerization synthesis used so far should be improved, in particular by reducing the molar mass dispersity or by synthesizing diblock copolymers where one block would contain the phosphonic acids and the other the PEG chains. Another promising approach is the synthesis of terpolymers on the same basis as for $MPEG_{ik}$-*co*-MPh in which amine-terminated PEG are appended on the same methacrylic acid backbone. This amine will then be used as a reactional group for covalent binding to add molecules of interest, such as fluorescent molecules for fluorescent imaging, a Gd-DOTA complex for an MRI contrast agent or targeting peptides for delivery. This last approach has been attempted recently and has shown promising results for $CeO_2$ particles used as antioxidant nanozymes in the treatment of stroke.[109] From the above outcomes, we conclude that this versatile and multi-substrate coating platform represents a step toward the understanding and control of nanomaterial interfacial properties. The summary of the knowledge obtained here is that biocompatible polymers can be pursued as surface modifiers of metal oxide nanoparticles and surfaces for a wide range of applications in material science and nanomedicine.

## ■ ACKNOWLEDGMENTS

We thank V. Baldim, N. Bia, Q. Crouzet, B.T. Doan, N. Giamblanco, A. Grein-Iankovski, W. Loh, C. Loubat, G. Marletta, N. Mignet, G. Ramniceanu, V. Torrisi, L. Vitorazi, C. Vezignol for their contributions to the research on phosphonic acid coatings. The coating platform described in this review would not have been possible without their commitment and expertise in the fields of chemistry, physics and biology. We are also indebted to the University of Paris (Dr. N. Mignet), University of Catania (Prof. G. Marletta) and to the University of Campinas (Prof. W. Loh) for their support during these studies, and to the access to their facilities.

## ■ FUNDING

ANR (Agence Nationale de la Recherche) and CGI (Commissariat à l'Investissement d'Avenir) are gratefully acknowledged for their financial support of this work through Labex SEAM (Science and Engineering for Advanced Materials and devices) ANR 11 LABX 086, ANR 11 IDEX 05 02. We acknowledge the ImagoSeine facility (Jacques Monod Institute, Paris, France), and the France BioImaging infrastructure supported by the French National Research Agency (ANR-10-INSB-04, « Investments for the future »). This research was supported in part by the Agence Nationale de la Recherche under the contract ANR-13-BS08-0015 (PANORAMA), ANR-12-CHEX-0011 (PULMONANO), ANR-15-CE18-0024-01 (ICONS), ANR-17-CE09-0017 (AlveolusMimics) and by Solvay.

## ■ ABBREVIATIONS

DMEM, Dulbecco's Modified Eagle's medium; EPR, Enhanced Permeability and Retention; ICP-OES, Inductively Coupled Plasma - Optical Emission Spectrometry; L-*b*-L, Layer-*by*-Layer; LCST, Low Critical Solution Temperature; MA, Methacrylate monomer bearing carboxylic acid functional group; MILC, Mass of metal Internalized/Adsorbed by Living Cells; MOx, Metal oxide; MPAm, Methacrylate monomer bearing a propylacrylamide group; MPAm-*co*-MPh, Statistical copolymer synthesized by free radical polymerization from MPAm and MPh monomers; $MPEG_{ik}$, PEG methacrylate macromonomer with PEG molecular weight of i kDa; $MPEG_{ik}$a, Amine-terminated PEG methacrylate macromonomer with PEG molecular weight of i kDa. $MPEG_{ik}$-*co*-MA, Statistical copolymer synthesized by free radical polymerization from $MPEG_{ik}$ and MA monomers; $MPEG_{ik}$-*co*-MPh, Statistical copolymer synthesized by free radical polymerization from $MPEG_{ik}$ and MPh monomers; $MPEG_{ik}$-*co*-$MPEG_{jk}$a-*co*-MPh, Statistical terpolymer synthesized by free radical polymerization from $MPEG_{ik}$, $MPEG_{jk}$a and MPh monomers; MPh, Methacrylic monomer bearing a phosphonic acid functional group; MRI, Magnetic Resonance Imaging; NP, Nanoparticle; OEG, Oligo(ethylene glycol); $OEG_{ik}$-Ph, Oligo(ethylene glycol) of molecular weight i kDa terminated with a single phosphonic acid; $PAA_{ik}$, Poly(acrylic acid) of molecular weight i kDa; PBS, Phosphate Buffered Saline; PEG, Poly(ethylene glycol); $PEG_{ik}$-Ph, Poly(ethylene glycol) of molecular weight terminated with a single phosphonic acid; PNIPAm, Poly(N-isopropylacrylamide); PNnPAm, Poly(N-n-propylacrylamide); QCM-D, Quartz Crystal Microbalance with Dissipation; RPMI, Roswell Park Memorial Institute medium; SAM, Self-Assembled Monolayer; TEM, Transmission Electron Microscopy

## ■ PHYSICAL QUANTITIES

$a$ (nm), Length of chemical monomer of a polymer chain; $b$ (nm), Kuhn length of a polymer chain; $c$ (wt. %, g $L^{-1}$), Solute weight concentration; $c_{NP}$ (wt. %, g $L^{-1}$), Nanoparticle weight concentration; $c_{Pol}$ (wt. %, g $L^{-1}$), Polymer weight concentration; $C$, Sauerbrey constant (17.7 ng s $cm^{-2}$); $D_H$ (nm), Hydrodynamic diameter; $f_0$ (Hz) , Resonance frequency in quartz crystal microbalance with dissipation experiment; $\Delta f_n$ (Hz), Resonance frequency shift observed after adsorption or desorption on QCM sensor, n being the overtone index; $h_{2D}$ (nm), Thickness of polymer adlayer on macroscopic flat substrate; $h_{NP}$ (nm), Thickness of polymer shell on nanoparticle; $I_{MRI}$, Magnetic resonence imaging intensity; $I_S$ (kcps), Scattering intensity from light scattering; $M_{Ce}$ (pg $cell^{-1}$), Mass of cerium adsorbed and internalized in cells; $M_{Fe}$ (pg $cell^{-1}$), Mass of iron adsorbed and internalized in cells; $\Delta m$ (ng $cm^{-2}$), Areal mass density determined in QCM-D; $N$, Degree of polymerization of a polymer; $n_{ads}$, Number of adsorbed chains per particle; $R_g$ (nm), Radius of gyration; $\sigma$ ($nm^{-2}$), Density of polymers adsorbed on a surface; $\sigma_{PEG}$ ($nm^{-2}$), Density of poly(ethylene glycol) adsorbed on a surface; $\Sigma$, Reduced tethered PEG density ($\Sigma = \pi \sigma_{PEG} R_g^2$);

$t_{1/2}^{clear}$ (min), Characteristic clearence time in organ pharmacokinetic monitoring; $t_{1/2}^{up}$ (min), Characteristic uptake time in organ pharmacokinetic monitoring; $X$, Mixing ratio; $X_C$, Critical mixing ratio.

## ■ TABLE OF CONTENTS GRAPHIC

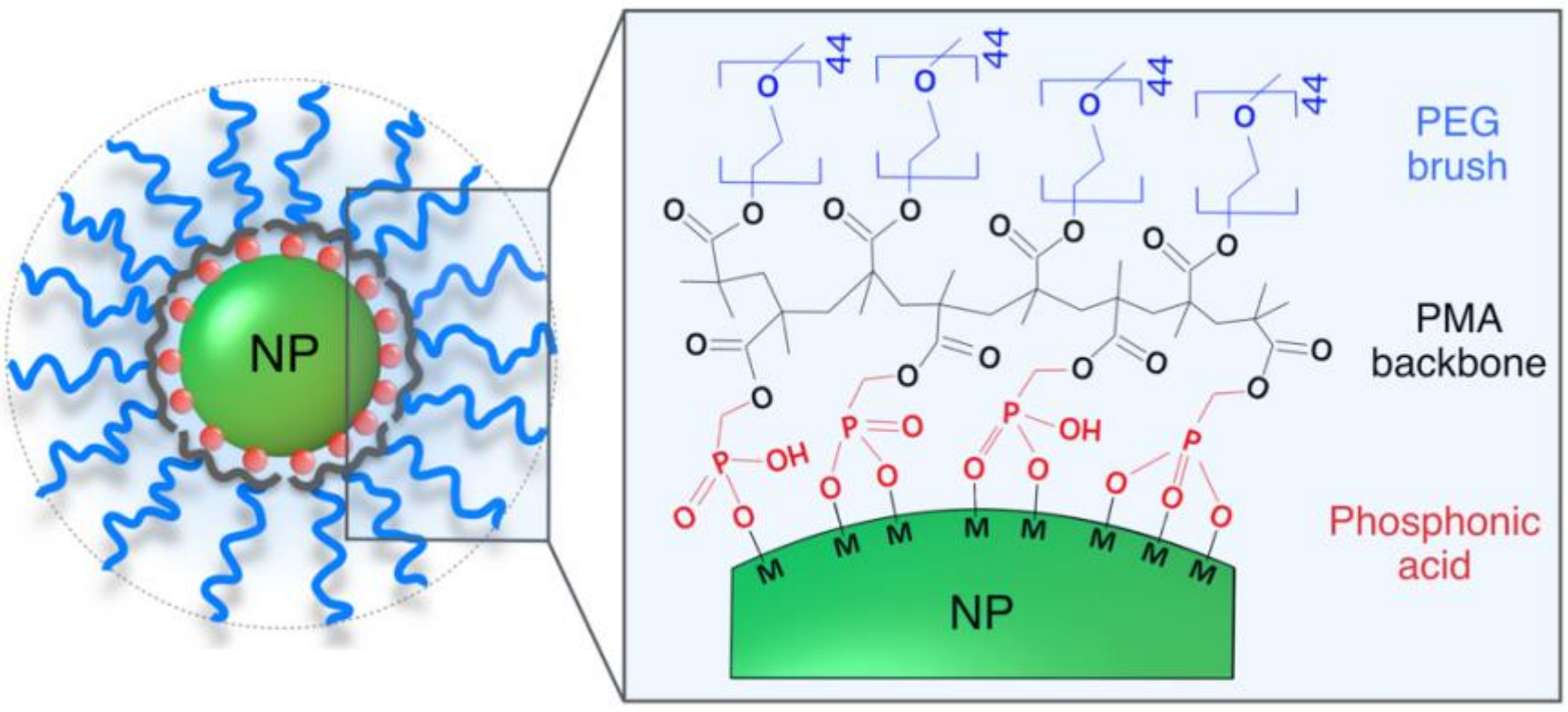

**TOC:** Schematic representation of a metal oxide nanoparticle coated with statistical phosphonic acid poly(ethylene glycol) copolymers using multidentate binding modes.

## ■ AUTHOR INFORMATION

**Corresponding author**
Jean-François Berret - *Université de Paris, CNRS, Matière et systèmes complexes, 75013 Paris, France*
https://orcid.org/0000-0001-5458-8653
Email: jean-francois.berret@u-paris.fr

**Author**
Alain Graillot - *Specific Polymers, ZAC Via Domitia, 150 Avenue des Cocardières, 34160 Castries, France.*
https://orcid.org/0000-0001-5970-5687
Email: alain.graillot@specificpolymers.fr

**Biographies**

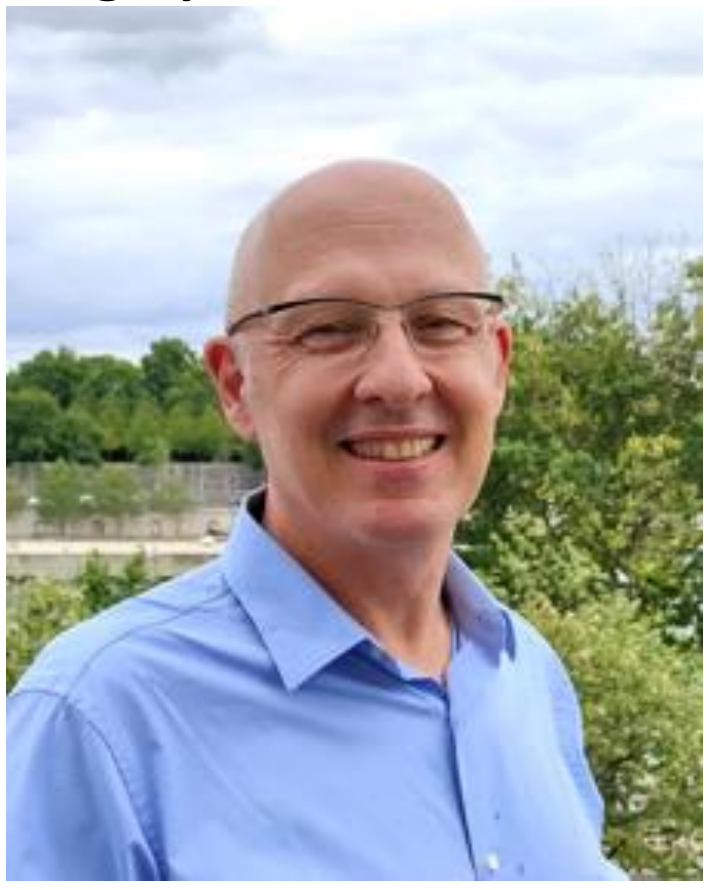

Jean-François Berret is director of research at the Centre National de la Recherche Scientifique, France. He is a renowned expert in soft condensed matter and biophysics, and a member of the Matière et Systèmes Complexes laboratory at the University of Paris. His current research focuses on the development of novel functional structures, devices, and systems with stimulus-response features at the nano- and microscale. His objectives are applications in the fields of medicine, biology and environment. Recently, he has been working on the synthesis and advanced coating of nanoceria for healthcare sciences.

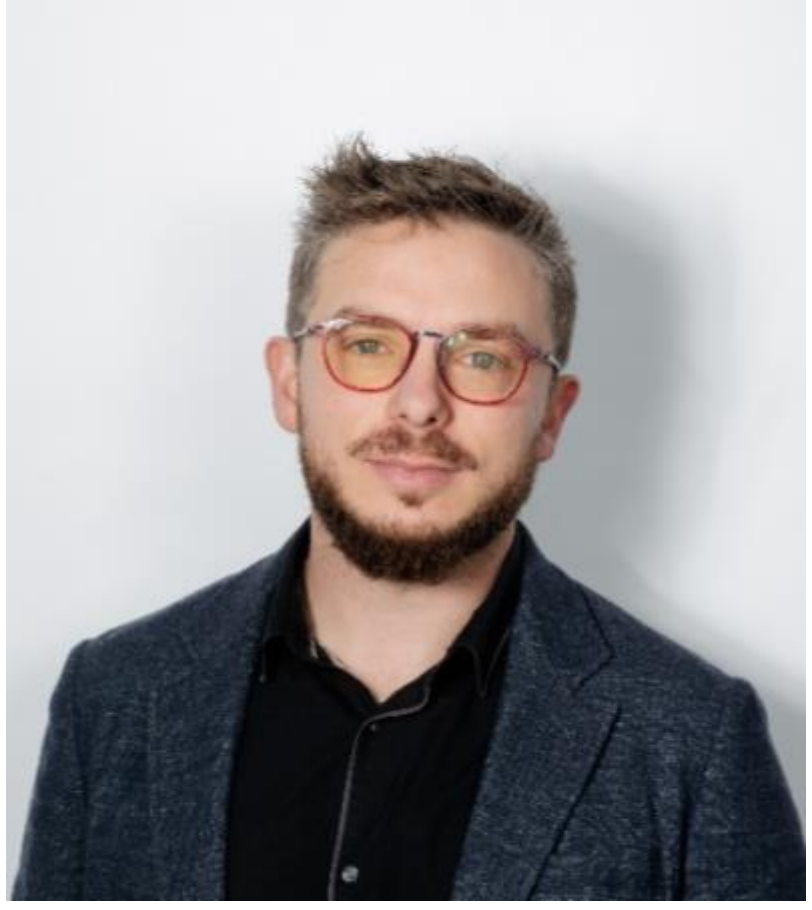

Alain Graillot is the Chief Operating Officer of SPECIFIC POLYMERS company, an SME providing R&D services in the field of innovative polymer and materials. He obtained his Ph.D. degree in polymer chemistry (RAFT controlled radical polymerization of phosphonated monomers) from Montpellier University in 2013. Within the company, he now leads multiple collaborative projects where he brings his expertise in various research areas: biomedical and nanomedicine, sustainable thermoset materials, polymeric materials for energy or high performance materials and coatings.

**Notes**

The authors declare no competing financial interest.

## ■ REFERENCES